\documentclass[a4paper,11pt]{article}
\pdfoutput=1 

\usepackage{jheppub} 

\usepackage{bm}
\usepackage{tikz}
\usepackage{multirow}
\usepackage{amsmath}
\usepackage{slashed}
\usepackage{braket}
\usepackage{amstext,amssymb}
\usepackage{graphicx}
\usepackage{booktabs}
\usepackage{IEEEtrantools}
\usepackage{hyperref}
\usepackage{xspace}
\usepackage{color}
\usepackage{units}
\usepackage{orcidlink}
\usepackage{xcolor}
\usepackage{verbatim}
\pagecolor{white}   

\title{Neutron star with dark matter using vector portal}
\author[a]{\hspace{-0.1cm}Deepak Kumar\orcidlink{0000-0001-9292-3598}} 
\author[b]{\hspace{-0.2cm}, Ranjita K. Mohapatra\orcidlink{0009-0003-4048-365X}}
\author[c,d]{\hspace{-0.2cm}, Hiranmaya Mishra\orcidlink{0000-0001-8128-1382}}
\author[e,c]{\hspace{-0.2cm}, Sudhanwa Patra\orcidlink{0000-0002-7469-2279}}

\affiliation[a]{Department of Physics, Indian Institute of Science Education and Research Berhampur, 760003, India}
\affiliation[b]{Department of Physics, Rajdhani College, Bhubaneswar 751003, India}
\affiliation[c]{Institute of Physics Bhubaneswar, Sachivalaya Marg, Bhubaneswar 751005, India}
\affiliation[d]{School of Physical Sciences, National Institute of Science Education and Research, An OCC of Homi Bhabha National Institute, Jatni - 752050, India}
\affiliation[e]{Department of Physics, Indian Institute of Technology Bhilai, Durg 491002, India}
\emailAdd{deepakk@iiserbpr.ac.in}
\emailAdd{ranjita.iop@gmail.com}
\emailAdd{hiranmaya@niser.ac.in}
\emailAdd{sudhanwa@iitbhilai.ac.in}

\abstract{
Compact astrophysical objects, such as neutron star, can provide a unique environment where the interplay between strongly interacting nuclear matter and dark matter (DM) can yield possible observable signatures. We investigate here the impact of fermionic DM interacting with nucleons via a vector mediator ($Z'$) portal inside neutron stars using the relativistic mean-field (RMF) framework. Unlike scalar portal DM models, which primarily modify the effective nucleon mass through scalar interactions, vector mediators ($Z'$) introduce additional repulsive interactions that directly affect the baryonic chemical potential and the pressure of dense matter. We show that the precise measurements of neutron star properties, including the mass–radius relation and tidal deformability from gravitational wave observations, X-ray and radio observations of pulsars, can shed light on properties of DM. We study the gross structural properties of a neutron star using the Tolman–Oppenheimer–Volkoff (TOV) equations, employing an equation of state (EOS) for neutron star matter in the presence of vector portal-assisted DM. The resulting stellar configurations consistent with observational bounds from gravitational wave observations (GW170817) in LIGO/Virgo and X-ray observations of pulsar PSR J0030+0451 in NICER, are shown to constrain the vector portal DM parameters. It is observed that, while large portal mass can soften the EOS of the DM admixed neutron star matter, the light portal mass can make the EOS stiffer at large densities resulting in distinct mass-radius relation and the tidal deformability between the two scenarios. The vector portal DM scenario, with DM interaction with quarks via $Z^\prime$ vector boson, can establish a direct connection to terrestrial searches, including direct and indirect detection and collider searches for the $Z'$ boson. Taken together, these constitute a comprehensive framework that bridges neutron star astrophysics with particle physics, enabling a multi-messenger exploration of DM properties.
}

\makeatletter \gdef\@fpheader{} \makeatother

\begin{document} 
\maketitle

\section{Introduction} \label{sec:introduction}
The dark matter (DM) is an enigmatic and invisible component of the Universe constituting about $85\%$ of the total matter content. Its existence has been primarily determined through a wide range of astrophysical and cosmological observations~\cite{Bertone:2004pz} including galactic rotation curves~\cite{Rubin:1970zza, Vanderheyden:2021gpj}, gravitational lensing~\cite{Israel:2016qsf,Clowe:2006eq}, large-scale structure formation and precision measurements of the cosmic microwave background~\cite{WMAP:2010qai}. Despite extensive efforts, the precise nature and composition of DM remains elusive. There are extensive and interesting experimental investigations that include the recent advancement in direct-detection experiments of DM such as LUX~\cite{daSilva:2017swg}, XENON-100~\cite{XENON100:2012itz}, PANDAX-II~\cite{PandaX-II:2016vec,PandaX-II:2017hlx}, XENON-1T \cite{XENON:2015gkh, XENON:2018voc}, and indirect-detection experiments such as PAMELA~\cite{PAMELA:2013vxg,PAMELA:2011bbe}, Fermi Gamma-ray space Telescope~\cite{Fermi-LAT:2009ihh} and IceCube~\cite{IceCube:2017rdn,IceCube:2018tkk} constraining various DM parameter space. However, there has been no hint of signatures for particle nature of DM. This motivates for complementary exploration for properties of DM through alternative astrophysical observations. In this context, the compact astrophysical objects such as NSs offer a powerful laboratory to probe DM interactions under extreme conditions of density and gravity. 

DM can influence neutron star (NS) modifying their mass-radius relations, alter tidal deformability and changing the cooling rate \cite{Barbat:2024yvi, Bramante:2023djs, Deliyergiyev:2019vti, Das:2018frc, Das:2020ecp, Ivanytskyi:2019wxd, LZ:2024psa}. The presence of DM inside NS has been studied in various context including asymmetric DM, bosonic condensates, fermionic DM, WIMPS, dark photons, axions etc, each leading to different astrophysical signatures~\cite{Bell:2013xk, Mukhopadhyay:2016dsg,Panotopoulos:2018ipq,Rutherford:2022xeb,Karkevandi:2024vov,Thakur:2023aqm,Mariani:2023wtv,Diedrichs:2023trk,Kouvaris:2007ay,Das:2018frc,Dutra:2022mxl,Lourenco:2022fmf,Flores:2024hts,Bertoni:2013bsa}. Axions, for example, as DM candidate, may address the cusp core problem in galactic halos and significantly influence NS cooling processes~\cite{Hinderer:2009ca, Postnikov:2010yn, Zacchi:2020dxl}; can potentially affect the stability of high mass NS~\cite{Lopes:2022efy} as well as lead to an increase the frequencies of the non-radial oscillations \cite{Kumar:2021hzo, Kumar:2025vku, Kumar:2024abb}. The NSs can be used for the detection of DM that may possibly reveal DM properties through their interactions with the nucleons within their cores \cite{Alexander:2020wpm, Kouvaris:2015rea}. DM may also influence NS heating and cooling using DM admixed equation of state (EOS) models showing their impact on thermal evolution and axion emission from proton NSs \cite{Hutauruk:2023nqc, Fischer:2021jfm, Bar:2019ifz, Zeng:2021moz, Gau:2023rct, Fortin:2018aom, Fortin:2018ehg, Lenzi:2022ypb,Kouvaris:2010vv,Bertoni:2013bsa}. 

There have been two main approaches for theoretically investigating the effects of DM on such NSs. It can be studied using the two fluid formalism where DM and ordinary matter are assumed to be interacting gravitationally. Consequently, the EOS for DM and ordinary matter are calculated separately with their coupling introduced gravitationally using the two fluid TOV framework \cite{Sotani:2025lzy, Issifu:2025jac, Pattnaik:2025edr, karan:2025kel, Routaray:2024fcq, Marzola:2024ame, Das:2020ecp, Harko:2011nu}. In the second approach i.e. a single fluid framework where non-gravitational interactions between DM and ordinary matter are assumed. Recently, there have been a large class of investigation that have focused on scalar portal DM models in which the DM interacts with nucleons via scalar mediators, often leading to modifications of the effective nucleon mass or additional attractive interactions in dense matter. These scalar interactions typically soften the EOS and can affect the mass-radius relation and tidal deformability of NSs \cite{Klangburam:2025rcb}. There have been approaches to investigate the impact of an axion like particles (ALPs) mediated DM interactions on NS properties \cite{Klangburam:2025rcb}.

Apart from scalar and pseudo-scalar portals, it may be noted that the portal between DM and ordinary matter can also be vector like. The vector portal DM models introduce interactions mediated by a new neutral gauge boson, commonly denoted as $Z'_\mu$~\cite{Taramati:2024kkn, Patel:2024zsu, Patra:2016ofq, Patra:2016shz}. In these models, DM fermion couples to $Z'$ boson, which in turn interacts with Standard Model (SM) quarks through vector currents. Such interactions naturally arise in extensions of the SM involving additional $U(1)$ gauge symmetries and have been extensively explored in the context of particle physics phenomenology. 

The vector portal scenario differs qualitatively from scalar portal models in several important ways. First, vector interactions modify the effective chemical potentials of nucleons and DM rather than the effective nucleon mass. This leads to an increase in the pressure of NS matter stiffening the EOS as compared to the scalar portal. Second, vector mediators provide a natural connection between NS physics and terrestrial experiments. The same $Z'$ mediator responsible for nucleon–DM interactions in NSs can lead to observable signatures in direct detection experiments through elastic DM–nucleon scattering, indirect detection via DM annihilation channels, and collider searches through dilepton or dijet resonance signatures.

In these contexts, the vector portal frameworks are particularly attractive for exploring DM properties using both terrestrial and astrophysical observations. It may be noted that the recent observations of NSs have resulted in reliable measurement of NS masses and constraints on their size. Radio observations of pulsars (PSR J1614–2230 \cite{Demorest:2010bx}, PSR J0348+0432 \cite{Antoniadis:2013pzd}, PSR J0740+6620 \cite{Salmi:2024aum, Miller:2021qha, Dittmann:2024mbo, Riley:2021pdl, Choudhury:2024xbk}) provide compelling evidence for NS of masses larger than $2M_\odot$ \cite{Demorest:2010bx, Antoniadis:2013pzd, Miller:2016kae, NANOGrav:2019jur, Ozel:2010bz}. On the other hand, gravitational wave observations (GW170817) in LIGO/Virgo~\cite{LIGOScientific:2017vwq, LIGOScientific:2017ync} and X-ray observations of pulsar PSR J0030+0451 in NICER \cite{Miller:2019cac, Vinciguerra:2023qxq} suggest that NSs with canonical mass of $1.4~M_\odot$ have radii in the range of $11-13\ {\rm km}$. Taken together, it means that the pressure of NS matter is small relatively upto twice nuclear matter density. Higher mass pulsar PSR J0740+6620 observations with radius $R = 12.39^{+1.30}_{-0.98}\, {\rm km}$ and mass $M = 2.08 \pm 0.07\, {\rm M}_{\odot}$ \cite{Miller:2021qha, Dittmann:2024mbo, Riley:2021pdl, Choudhury:2024xbk, Salmi:2024aum} suggests high pressure at the core to support such high mass NS.

Quantum Hadrodynamics (QHD) offers a well-established theoretical framework to describe nuclear interactions in NSs within the relativistic mean-field (RMF) approach. In this formalism, nucleons interact through the exchange of mesonic fields, primarily a scalar meson describing attraction and a vector meson responsible for repulsion among nucleons. The RMF-based QHD model successfully reproduces key nuclear saturation properties as well as the structure of finite nuclei. It also serves as a foundation for many NS EOSs. By extending this framework to include dark matter interactions mediated by a vector boson, one can systematically explore the impact of such additional interactions on the nuclear EOS and their effects on the resulting NS observables.

With these motivations, we investigate here a fermionic DM interacting with nucleonic matter inside NSs through a vector mediator portal. We employ the RMF approach to describe nuclear interactions and extend it by including a new $U(1)_X$ gauge boson that couples both to DM and to quarks thereby to nucleons. Within this novel vector portal framework, we perform a detailed study of NS observables enabling to a comprehensive exploration of vector portal DM. We shall also look into  the constraints arising from NS observables on the parameters of DM and the portal which can be complementary to terrestrial experimental constraints for the same. 

The paper is structured as follows --- In section \ref{sec:formalism}, we present the theoretical model including RMF formalism for nuclear matter and the vector boson mediated interactions with fermionic DM. In section \ref{sec:neutron_star_structure}, we give the necessary formalism that is used to solve for the NS structure and tidal deformability. In section \ref{sec:results_and_discussion}, we discuss the results of the present analysis including the effect of vector boson portal on DM EoS, mass-radius relation and tidal deformability. Finally, in section \ref{sec:summary_and_conclusion} we summarize our findings outlining potential directions for further investigation.

\section{The vector portal fermionic DM model} \label{sec:formalism}
The RMF theory has been a powerful tool in describing hadronic matter under extreme conditions. Of late, the impact of DM on NS properties has attracted significant interest. In the present investigation, we make use of the nucleonic RMF model and extend it by including interactions between nucleons and DM particles via a vector boson $Z^\prime_\mu$ portal. In this scenario, a new $U(1)_X$ gauge boson ($Z^\prime_\mu$) mediates interactions between the dark sector and the SM quarks. The $Z'$ couples both to the DM fermion $\chi$ and to the quark current, thereby inducing an effective nucleon--DM interaction inside dense nuclear matter. 

In the RMF framework, the nucleon meson interaction is described by the Lagrangian density~ \cite{Mishra:2001py, Tolos:2016hhl}
\begin{align}
\mathcal{L}_{\rm HM} &= \sum_{\alpha=p,n} \bar{\psi}_{\alpha}
    \Big[\gamma_{\mu}\big(i\partial^{\mu}-g_{\omega {\alpha}}\omega^{\mu}
    -\tfrac{1}{2}g_{\rho {\alpha}}\vec{\tau}\!\cdot\!\vec{\rho}^{\,\mu}
    \big)
    - (M - g_{\sigma {\alpha}} \sigma) \Big]\psi_\alpha \nonumber \\
    & \quad + \tfrac{1}{2} \big( \partial_\mu \sigma \partial^\mu \sigma - m_\sigma^2 \sigma^2 \big) + V_{\rm NL} \nonumber \\
    & \quad - \tfrac{1}{4}\Omega_{\mu\nu}\Omega^{\mu\nu} + \tfrac{1}{2}m_\omega^2 \omega_\mu \omega^\mu
    - \tfrac{1}{4}\, \vec{R}_{\mu\nu} \cdot \vec{R}^{\mu\nu} + \tfrac{1}{2}m_\rho^2\, \vec{\rho}_\mu \cdot \vec{\rho}^{\,\mu},
\label{eq:L_HAD_VEC}
\end{align}
\begin{IEEEeqnarray}{rCl}
\hspace{-0.5cm}\mbox{with,\,} \quad V_{\rm NL} &=& \frac{\kappa}{3!}(g_{\sigma {\rm N}}\sigma)^3 + \frac{\lambda}{4!}(g_{\sigma {\rm N}}\sigma)^4 - \frac{\xi_\omega}{4!}(g_{\omega {\rm N}}^2\omega_{\mu}\omega^{\mu})^2 - {\Lambda_{\omega\rho}}
\big(g_{\omega {\rm N}} ^{2} \omega_{\mu} \omega^{\mu}\big)
\big(
g_{\rho {\rm N}}^2\vec{\rho}_{\mu} \cdot \vec{\rho}^{\mu} \big).
\end{IEEEeqnarray}
Here, $\psi_{\alpha}$ ($\alpha = p,n$) represents the nucleon field, $\sigma$ is the scalar meson field mediating an attractive interaction, $\omega_\mu$ is the vector meson field mediating the repulsive interactions and $\vec{\rho}_\mu$ is the isovector meson field accounting for isospin asymmetry. The non-linear scalar self-interactions involving $\kappa$ and $\lambda$ are included to reproduce the empirical properties of nucleonic matter~\cite{Walecka:1974ef, Boguta:1977xi, Boguta:1983sm, Serot:1997xg} and other non-linear terms are also included to satisfy higher order saturation properties of dense nuclear matter at saturation density. The field strength tensors for vector mesons are defined as $\Omega_{\mu \nu} = \partial_{\mu}\omega_{\nu} - \partial_{\nu}\omega_{\mu}$, and $\vec{R}_{\mu \nu} = \partial_{\mu}\vec{\rho}_{\nu} - \partial_{\nu}\vec{\rho}_{\mu}$. The meson-baryon couplings $g_{\sigma}$, $g_{\omega}$ and $g_{\rho}$ are denoted for the scalar, vector and isovector coupling constants, respectively. 

To incorporate DM interactions, we extend the Lagrangian by introducing fermionic DM that interacts with SM via a $Z'_\mu$ portal as follows,
\begin{align}
\mathcal{L}_{\rm DM} &= \bar{\chi}(i\gamma^\mu \partial_\mu - M_{\chi})\chi - g_{\chi Z^\prime} \bar{\chi}\gamma^\mu \chi Z'_\mu  - \sum_{q=u,d} g_q\, \bar{q}\gamma^\mu q \, Z'_\mu - \tfrac{1}{4} Z'_{\mu\nu}Z'^{\mu\nu} + \tfrac{1}{2} m_{Z'}^2 Z'_\mu Z'^\mu, \label{eq:L_DM_VEC}
\end{align}
The $\chi$ represents fermionic DM field of mass $M_{\chi}$ which interacts with the vector boson $Z^\prime_\mu$. Here, $g_\chi$ and $g_q$ denote the DM--$Z'$ and quark--$Z'$ couplings, respectively. The effective nucleon-$Z^\prime$ Lagrangian at low energy ($q^2 << \Lambda^2_{\rm QCD}$) is obtained by evaluating the matrix element of the quark current $(\bar{q} \gamma^\mu q)$ between nucleonic states. The resulting effective coupling of $Z^\prime$ with nucleus is given by
\begin{eqnarray}
\mathcal{L}_{Z^\prime \psi} = Z^\prime_\mu \sum_{\alpha =p,n} g_{\alpha Z^\prime} \bar{\psi}_{\alpha} \gamma^\mu \psi_{\alpha}
\end{eqnarray}
with $g_{\alpha Z^\prime} \simeq 3 g_q$ with $g_{p Z^\prime} = g_{n Z^\prime} \equiv g_{N Z^\prime}  = 3g_q$. Thus, the interaction of $Z^\prime$ with nucleons is primarily vector like resulting in a repulsive potential between the nucleons ~\cite{Bishara:2017pfq}. {The RMF model parameters used here are the Bayesian improved \cite{Kumar:2026tqe} which are presented  in Table \ref{tab:hadronic_matter_model_params}. These parameters  saturate the nuclear matter at $\rho_0 \simeq 0.15\,\text{fm}^{-3}$ with a binding energy of $-16.0$~MeV and supports a maximum NS mass more than $2M_{\odot}$, aligning with the current observational constraints}. Although the present RMF model with $\sigma$, $\omega$, $\rho$ mesons captures the essential physics of nucleonic matter with vector portal DM, there are richer RMF frameworks, such as those incorporating additional meson fields or hyperons. Similarly, alternative dark sector models with different DM interactions involving scalar and pseudo-scalar portal have been attempted yielding distinct effects. The present investigation is a minimal yet novel extension of the RMF model to explore DM-nucleon coupling via vector portal serving as a baseline for future comparisons.

The mean-field approximations correspond to taking the meson fields as classical while retaining the quantum nature of fermionic fields. The meson fields (vector bosons)  are expressed by their expectation values in the medium of baryonic matter (denoted by subscript 0) $i.e.$ $\langle\sigma\rangle = \sigma_0$, $\langle \omega_\mu\rangle=\omega_0\delta_{\mu 0}$, $\langle \rho_\mu^a\rangle$ =$\delta_{\mu 0}\delta_{3}^a \rho_{0}^{3}$, $\langle Z^\prime_\mu \rangle = Z^\prime_{0} \delta_{\mu 0}$. One can derive the Dirac equation for the nucleons and the DM as, 
\begin{align}
\big[i\gamma^\mu \partial_\mu - g_{\omega \alpha} \gamma^0 \omega_0 - g_{\rho \alpha} \gamma^0 \tau_3 \rho^{3}_0  - g_{N Z^\prime} \gamma^0 Z'_0 - M^* \big]\psi_{\alpha} &= 0, \\
\big[i\gamma^\mu \partial_\mu - g_{\chi Z^\prime} \gamma^0 Z'_0 - M_{\chi} \big]\chi &= 0,
\end{align}
with $M^*_{\alpha} = M_{\alpha} - g_{\sigma\alpha} \sigma_0$.

The equations of motion of meson fields can be found by the Euler-Lagrange equations for the meson fields using the Lagrangian 
\begin{IEEEeqnarray}{rCl}
m_{\sigma}^2 \sigma_0 &=& \sum_{\alpha=p,n} g_{\sigma \alpha}n_{\alpha}^s - \frac{\kappa}{2} g_{\sigma {\rm N}}^3 \sigma_0^2 - \frac{\lambda}{3!} g_{\sigma {\rm N}}^4 \sigma_0^3, \label{fieldeqns.sigma} \\
m_{\omega}^2 \omega_0 &=& \sum_{\alpha=p,n} g_{\omega \alpha}n_{\alpha} 
- \frac{\xi_\omega}{3!} g_{\omega {\rm N}}^4 \omega_0^3 - 2\Lambda_{\omega\rho} (g_{\rho {\rm N}} g_{\omega {\rm N}}\rho_0)^2 \omega_0, \label{fieldeqns.omega} \\
m_{\rho}^2 \rho_3^0 &=& \sum_{\alpha=p,n} g_{\rho}I_{3\alpha}n_{\alpha} - 2\Lambda_{\omega\rho} (g_{\rho {\rm N}} g_{\omega {\rm N}}\omega_0)^2 \rho_0, \label{fieldeqns.rho} \\
{m_{Z'}^2 Z'_0} &=& {g_{\chi Z^{\prime}}n_\chi  + \sum_{\alpha=p,n} g_{N Z^{\prime}} n_{\alpha}}
\label{fieldeqns.higgs}
\end{IEEEeqnarray}
where, $I_{3\alpha}$ is the third component of the isospin of a $\alpha^{\rm th}$ baryon. We have taken $I_{3 (p,n)} = \left(\frac{1}{2}, -\frac{1}{2}\right)$. The baryon and DM number densities are given by
\begin{IEEEeqnarray}{rCl}
n_B &:=& \sum_{\alpha=p,n} \langle\psi_{\alpha}^{\dagger}\psi_{\alpha}\rangle = \sum_{\alpha=p,n} \gamma_{\alpha} \frac{k_{F\alpha}^3}{6\pi^2}, \label{baryon.density}
\\
n_{\alpha}^s &:=& \langle\bar{\psi}_{\alpha}\psi_{\alpha}\rangle = \gamma_{\alpha} \int_0^{k_{F\alpha}} \frac{d^3k}{(2\pi)^3} \frac{M_{\alpha}^*}{\sqrt{M_{\alpha}^{*}{^2} + k^2}} , \label{baryon.scalar.density}
\\
n_{\chi} &:=& \langle \chi^\dagger\chi\rangle = \gamma_{\chi} \frac{k_{F \chi}^3}{6\pi^2}, \label{dark.matter.scalar.density}
\end{IEEEeqnarray}
Where $\gamma_{\alpha (\chi)} = 2$ is the spin degeneracy factor for fermions and $k_{F\alpha}$, $k_{F\chi}$ are the Fermi momenta of the nucleons and DM particles, respectively. The Fermi moment of the nucleon $k_{F\alpha}$ is given as follows,
\begin{equation}
k_{F\alpha}=\sqrt{\mu_{\alpha}^*{}^2-M_{\alpha}^*{}^2} \iff \mu_{\alpha}^* > M_{\alpha}^* \quad {\rm otherwise} \quad k_{F\alpha} = 0
\end{equation}
with an effective baryonic chemical potential, $\mu_{\alpha}^*$ given as
\begin{equation}
\mu_{\alpha}^* = \mu_{\alpha} - g_{\omega\alpha}\omega_0 - g_{\rho}I_{3\alpha}\rho_0^3 - g_{N Z^\prime} Z^\prime_0, \label{effective-chemical-potential-nl3}
\end{equation}
where ${\mu}_{\alpha} = {\mu}_B + q_{\alpha} {\mu}_E$. Here ${\mu}_B$ and ${\mu}_E$ are baryon chemical potential and electric chemical potential respectively. Similarly, for DM, $k_{F\chi} = \sqrt{\mu_\chi^*{}^2 - M_{\chi}^2} \iff \mu_{\chi}^* > M_{\chi} \quad {\rm otherwise} \quad k_{F\chi} = 0$. The effective chemical potential of DM is defined as {$\mu_\chi^* = \mu_\chi - g_{\chi Z^{\prime}} Z_0^{\prime}$}.

The NSs are globally charge neutral and the matter inside the core is under $\beta$-equilibrium ($p + e \to n$ and $p + \mu \to n$). This needs to include leptons i.e. electrons ($e$) and muons ($\mu$) for the charge neutral NS matter. Thus, the chemical potentials and the number densities of the constituents of NS matter are related by the following equations,
\begin{IEEEeqnarray}{rCl}
{\mu}_{n} = {\mu}_p &+& {\mu}_e, \quad {\mu}_{n} = {\mu}_p + {\mu}_{\mu},  \label{beta.equalibrium.rmf}
\\
\sum_{\alpha = p,n,e,\mu} n_{\alpha} q_{\alpha} &=& 0,
\end{IEEEeqnarray}

With all these ingredients, we can obtain the total energy density, $\mathcal{E}_{\rm TOT} $, and $\mathcal{P}_{\rm TOT} $ and hence, the EOS of NS matter. The EOS of the system is calculated by summing the contributions from nucleons, leptons, DM, mesons and the vector bosons. The energy density is given by~
\begin{IEEEeqnarray}{rCl}
&&\hspace*{-1cm}\mathcal{E}_{\rm TOT} = \mathcal{E}_{\rm Baryons} + \mathcal{E}_{\rm \chi}+\mathcal{E}_{\rm Mesons}
+ \mathcal{E}_{Z^\prime}
+ \mathcal{E}_{\rm Lepton} \label{energy_density_nm}
 \\
&&\hspace{0.5cm} \mathcal{E}_{\rm Baryons} = 
\sum_{\alpha=p,n} \frac{\gamma_{\alpha}}{(2 \pi)^3}\int^{k_{F\alpha}}_{0}\, d^3k \sqrt{k^2 + {M_{\alpha}^*}^2 }
\equiv 
\sum_{\alpha=p,n} \frac{{M_{\alpha}^*}^4}{\pi^2} H(k_{F\alpha}/M_{\alpha}^*)
\\
&&\hspace{0.5cm} \mathcal{E}_{\chi} =  \frac{\gamma}{(2 \pi)^3}\int^{k_{F\chi}}_{0}\, d^3k \sqrt{k^2 + M_{\chi}^2 }
=
\frac{M_{\chi}^4}{\pi^2} H(k_{F\chi}/M_{\chi}) 
\\
&&\hspace{0.5cm} \mathcal{E}_{\rm Mesons} = 
\frac{1}{2}m_{\sigma}^2\sigma_0^2 
+ \frac{1}{2} m_{\omega}^2\omega_0^2 + \frac{1}{2} m_{\rho}^2{\rho_{0}^3}\,{}^2 
+  \frac{\kappa}{3!}(g_{\sigma{\rm N}}\sigma_0)^3 + \frac{\lambda}{4!}(g_{\sigma{\rm N}}\sigma_0)^4 \nonumber \\
&&\hspace{2cm}+ \frac{\xi_{\omega}}{8}(g_{\omega{\rm N}}\omega)^4 + 3{\Lambda_{\omega\rho}}(g_{\rho{\rm N}}g_{\omega{\rm N}}\rho_0 \omega_0)^2 
\\
&&\hspace{0.5cm} \mathcal{E}_{Z^\prime } =
\frac{1}{2} m^2_{Z^\prime} Z^2_0
\\ 
&&\hspace{0.5cm} \mathcal{E}_{\rm Lepton} = 
\sum_{\ell=e,\mu}\frac{m_\ell^4}{\pi^2} H(k_{F\ell}/m_l)  
\end{IEEEeqnarray}
Where we have introduced the function $H(z)$ which is given as
\begin{IEEEeqnarray}{rCl}
H(z) &=& \dfrac{1}{8} \left[z\sqrt{1+z^2}(1+2z^2)-\sinh^{-1}z \right], \label{function_h}
\end{IEEEeqnarray}

\noindent The total  pressure, $\mathcal{P}_{\rm TOT} $, can be found using the thermodynamic relation as
\begin{IEEEeqnarray}{rCl}
\mathcal{P}_{\rm TOT}  &=& \sum_{i=n,p,\ell,\chi} \mu_i n_i - \mathcal{E}_{\rm TOT}. \label{pressure_nm}
\end{IEEEeqnarray}
Thus, the effect of introducing the vector boson $Z^\prime$ lies in reducing the effective chemical potential as in Eq. (\ref{effective-chemical-potential-nl3}).

\section{Neutron star structure and its tidal deformability}
\label{sec:neutron_star_structure}
In this section, we describe the formalism that we use to study the properties of the NS. The metric for a static, spherically symmetric star, is given by \cite{Weinberg:1972kfs}
\begin{eqnarray}
ds^2 =  e^{2\nu(r)} dt^2 - e^{2\lambda(r)} dr^2 - r^2 \big(d\theta^2+\sin^2\theta d\phi^2 \big)\,,
\label{eq:metric}
\end{eqnarray}
where $\nu(r)$ and $\lambda(r)$ are the metric functions. It is convenient to define the mass function, $m(r)$ in favor of $\lambda(r)$ as 
\begin{equation}
e^{2 \lambda(r)} = \bigg(1-\frac{2 m(r)}{r} \bigg)^{-1}  
\end{equation}
Starting from the line element given in Eq.(\ref{eq:metric}), the equations for the structure of a relativistic spherical and static star composed of a perfect fluid were derived from Einstein's equation by Tolman–Oppenheimer–Volkoff  known as TOV equations \cite{Oppenheimer:1939ne,PhysRev.55.364},
\begin{eqnarray}
&&\frac{d \mathcal{P}(r)}{dr} =-  \frac{\big[{\mathcal E} +\mathcal{P}\big]\big[m + 4\pi r^3 \mathcal{P}\big]}{r(r-2m)}, \label{tov_pressure} \\
&&\frac{d m(r)}{dr}= 4\pi r^2 \mathcal{E} \label{tov_mass}
\end{eqnarray}
The above set of equations $\mathcal{E}(r)$, $\mathcal{P}(r)$, $m(r)$ are the energy densities, the pressure and the mass of the star enclosed within a radius $r$, respectively. The boundary conditions $m(r=0) = 0$; $\mathcal{P}(r=0)= \mathcal{P}_c$ and $\mathcal{P}(r=R)=0$ where $\mathcal{P}_c$ is the central pressure lead to equilibrium configurations in combination with EOS of NS matter, thus obtaining radius $R$ and mass $M=m(R)$ of NS for a given central pressure $\mathcal{P}_c$ or energy density $\mathcal{E}_c$. For a set of central densities $\mathcal{E}_c$, one can obtain the mass-radius (M-R) curve. 

The tidal distortion of NS in a binary system links the equation of state to the gravitational wave emission during the inspiral. The tidal deformability parameter quantifies the quadropole deformation of a compact object in a binary system due to the tidal effect of its companion star. The relation between the induced quadropole moment tensor and the tidal field tensor in leading order is given by, $Q_{ij} = - \lambda \mathcal{E}_{ij}$ where $\lambda$ is related to the tidal love number ($\ell=2$) \cite{Hinderer:2007mb}. The tidal love number as $k_2=3/2\, \lambda R^{-5}$, $R$ being the radius of the NS. One can estimate $k_2$ perturbatively by calculating the deformation $h_{\alpha \beta}$ of the metric from the spherically symmetric metric. The deformation of the metric in Regge-Wheler gauge can be written as \cite{Hinderer:2007mb}
\begin{equation}
h_{\alpha \beta } = \text{Diag}\Bigg[-
e^{2 \nu(r)} H_{0}(r),\ e^{2 \lambda(r)} H_{2}(r),\ r^2 K(r),\ r^2 \sin^2\theta K(r) 
\bigg] Y_{20} (\theta, \phi)
\label{eq:hegge}
\end{equation}
where $H_{0}$, $H_{2}$ and $K(r)$ are perturbed metric functions. It turns out that $H_{2}(r)=-H_{0}(r) \equiv H(r)$ using Einstein's equation 
$\delta g_{\alpha \beta} = \delta T_{\alpha \beta}$ while $K^\prime(r) = 2 H(r) \nu(r)$. The logarithm derivative of the deformation function $H(r)$ i.e, $y(r) = r\, \frac{H^\prime_0(r)}{H_0(r)}$ satisfies the first order differential equation \cite{PhysRevD.80.084035}
\begin{equation}
r\,y^\prime(r) + y(r)^2 + y(r) F(r) + r^2 Q(r) = 0\,. 
\label{tidal_y}
\end{equation}
Where the function $F(r)$, $Q(r)$ are given by
\begin{eqnarray}
   &&F(r) =  \big[1+ 4\pi r^2 \big(\mathcal{P} -\mathcal{E} \big) \big]
   \bigg(1 -\frac{2 M}{r} \bigg)^{-1}\,, \nonumber \\
   &&Q(r) = 4\pi 
   \bigg[ 5 \mathcal{E} + 9 \mathcal{P} + \frac{\mathcal{E}+\mathcal{P}}{d\mathcal{P}/d\mathcal{E}}
   \bigg] \bigg(1 -\frac{2 M}{r} \bigg)^{-1}
   - \frac{6}{r^2}\bigg(1 -\frac{2 M}{r} \bigg)^{-1}
   \nonumber \\
   &&\hspace{5cm}-\frac{4 M^2}{r^4} \bigg(1+ \frac{4\pi r^3 \mathcal{P}}{M} \bigg)^2 \bigg(1 -\frac{2 M}{r} \bigg)^{-2}
\end{eqnarray}
To calculate the tidal deformation, the equation for the metric perturbation given in Eq.(\ref{tidal_y}) can be integrated together with TOV Eqs.(\ref{tov_pressure},\ref{tov_mass}) for a given EOS radially outwards with the boundary conditions, 
$y(r=0)=2, \mathcal{P}(r=0) = \mathcal{P}_c$ and $M(r=0) = 0$. 

The tidal lover number $k_2$ is related to $y_{R} \equiv y(R)$ through 
\begin{eqnarray}
    k_2 &=& \frac{8 C^5}{5}\big( 1 -2 C^2 
    \big) \big[2 + 2C (y_R-1)-y_R \big] \times \nonumber \\
  &&\bigg\{2C(6-3 y_R+3C(5 y_R-8))
  + 4 C^3 \bigg[13 - 11 y_R + C (3 y_R-2) + 2 C^2 (1+y_R) \bigg] \nonumber \\
  &&
  + 3 (1-2C)^2 \bigg[2-y_R+2C(y_R-1) \bigg]
  \text{log}(1-2C)
  \bigg\}^{-1}  \label{love_number_k2}
\end{eqnarray}
where $C\equiv (M/R)$ is the compactness parameter of the star of mass $M$ and radius $R$. The dimensionless tidal deformability $\Lambda$ is defined as \cite{PhysRevD.77.021502,Hinderer:2007mb,PhysRevD.81.123016,PhysRevD.85.123007}
\begin{equation}
\Lambda = \frac{\lambda}{M^5} = \frac{2 k_2}{3 C^5}\, .\label{eq:tidal}
\end{equation}
The observable signature of relativistic tidal deformation will have an effect on the phase evolution of the gravitational wave spectrum from inspiral binary NS system. This signal will have cumulative effects of the tidal deformation arising from both the stars. Therefore, one can combine the tidal deformabilities and define a dimensionless tidal deformability taking a weighted average as 
\cite{PhysRevD.85.123007}
\begin{equation}
    \widetilde{\Lambda} = 
    \frac{16}{13}
    \bigg[\frac{ (M_1+ 12 M_2) M_1^4 \Lambda_1 
    + (M_2+ 12 M_1) M_1^4 \Lambda_2}
    {(M_1+M_2)^{5}}
    \bigg]
    \label{eq:tidal-final}
\end{equation}
In the above, $\Lambda_1$ and $\Lambda_2$ are the individual tidal deformabilities corresponding to the two components of NS binary with masses $M_1$ and $M_2$, respectively. 

\section{Results and discussion} \label{sec:results_and_discussion}
Next we shall discuss the numerical results regarding the impact of fermionic DM on NS properties within the RMF framework extended by a vector portal interaction mediated by a $Z'$ boson. As mentioned earlier, the vector portal introduces an additional repulsive interaction between DM and nuclear matter, whose strength depends on the mediator mass and the corresponding coupling constants. Regarding the parameters for the vector portal DM models in NSs we discuss three sets which are given in Table \ref{tab:dark_matter_model_params} ~ \cite{Taramati:2024kkn, Patel:2024zsu, Patra:2016ofq, Patra:2016shz}. {For the parameters given in sets-1 and set-2,  the coupling of quarks with $Z^\prime$ (equivalently, to nucleons, i.e, $g_{q Z'} \simeq 1/3\,g_{N Z'}$) is taken to be the same as the coupling $g_{\chi Z'}$ of the DM with the vector boson $Z^\prime$~\cite{Bishara:2017pfq, Borah:2025cqj}}. In set 1, we have taken a heavy $Z^\prime$ with a mass around $1800$~GeV along with the DM mass $m_\chi \simeq 200$~GeV.  Set 2 corresponds to a heavier DM mass $m_\chi \simeq 1800$~GeV and $m_{Z^\prime} \simeq \mbox{900}$~GeV. These are the limiting cases for parameters satisfying relic density, direct detection, and dijet bounds from collider studies~\cite{Taramati:2024kkn}. 

Unlike set-1 and set-2, which correspond to heavy vector portal mediators, the parameters of set-3 are motivated by self-interacting dark matter (SIDM) scenarios with a light vector mediator~\cite{Spergel:1999mh, Tulin:2013teo, Patel:2022qvv, Kouvaris:2014uoa, Bernal:2015ova, Kainulainen:2015sva, Hambye:2019tjt, Cirelli:2016rnw, Kahlhoefer:2017umn, Dutta:2021wbn}. Dark matter self-interactions can be mediated by either scalar or vector particles; however, vector mediators arise naturally in gauge extensions of the Standard Model and therefore provide a particularly well-motivated framework. In recent studies, for example in Refs.~\cite{Borah:2021pet,Patel:2022qvv}, the Standard Model is extended by an additional $U(1)_X$ gauge symmetry containing a vector-like fermionic dark matter candidate charged under the new gauge group. In such models, the $Z^\prime$ boson acts as the mediator responsible for dark matter self-interactions. Several other SIDM scenarios involving light vector mediators have also been studied in the literature \cite{KumarBarman:2018hla, Kamada:2018kmi, Balducci:2018ryj, Kamada:2018zxi, Lee:2020eap, Ho:2022erb, Heeck:2022znj}, which are relevant to the present investigation.

In the present work, we consider a dark matter particle with mass in the few-GeV range interacting through a light vector mediator with mass $m_{Z^\prime}\sim100$ MeV. Motivated by SIDM models, we adopt a hierarchical coupling structure in which the mediator couples strongly to the dark sector while maintaining highly suppressed couplings to visible-sector particles. In such scenarios, one may naturally consider couplings of the order $g_{\chi Z^\prime}=\mathcal{O}(0.1-1)$ together with $g_{qZ^\prime}\sim10^{-5}-10^{-3}$, which are consistent with current laboratory, beam-dump, meson-decay, and collider constraints on light vector mediators. In particular, we choose $g_{\chi Z^\prime}=0.8$ and $g_{qZ^\prime}=5\times10^{-4}$ for the benchmark considered in this work.

It is worth noting that this region of parameter space has been extensively investigated in SIDM models as a possible solution to several small-scale structure anomalies of collisionless cold dark matter, including the core-cusp problem~\cite{Navarro:1996gj}, the too-big-to-fail problem~\cite{Boylan-Kolchin:2011qkt,Boylan-Kolchin:2011lmk}, and the missing-satellites problem~\cite{Klypin:1999uc,Kauffmann:1993gv,Moore:1999nt}. A light mediator naturally generates velocity-dependent dark matter self-interactions that can yield the required self-scattering cross sections on galactic scales while remaining compatible with observations of galaxy clusters~\cite{Bahcall:1999xn,Springel:2006vs,Trujillo-Gomez:2010jbn}. Furthermore, the suppressed visible-sector coupling significantly weakens direct-detection and collider constraints, while the observed relic abundance can be achieved through dark-sector annihilation channels involving the light mediator~\cite{Tulin:2013teo,Borah:2021pet,Patel:2022qvv}. Consequently, the set-3 benchmark provides a phenomenologically viable realization of the vector portal framework and serves as an illustrative example for studying the impact of a light mediator on the equation of state and global properties of dark-matter-admixed neutron stars.

We next introduce another benchmark scenario, namely set-4, corresponding to a light-mediator strong-coupling benchmark designed to explore the maximal influence of a light vector mediator on the equation of state and structural properties of dark-matter-admixed neutron stars. In this benchmark, the mediator mass is again chosen in the light regime, $m_{Z^\prime}\sim100$ MeV, together with relatively large portal couplings to both the dark and visible sectors. As a consequence, the repulsive vector interaction can significantly modify the dark matter pressure and energy density, thereby producing a pronounced effect on the stiffness of the equation of state and the resulting mass-radius relation.

It should be emphasized, however, that such a parameter choice is generally difficult to accommodate within a minimal kinetically mixed dark-photon framework, where collider and direct-detection constraints typically require much smaller couplings to Standard Model particles. On the other hand, more general scenarios involving asymmetric dark matter, secluded dark sectors, or non-minimal gauge structures can substantially relax conventional terrestrial constraints because the relic abundance is generated through a primordial asymmetry and the dominant annihilation channels occur primarily within the dark sector~\cite{Kaplan:2009ag,Nussinov:1985xr,Kaplan:1991ah,Petraki:2013wwa,Zurek:2013wia,Tulin:2017ara,Kaplinghat:2015aga,Hooper:2004dc,Farrar:2005zd,Patel:2022xyv}. In such frameworks, larger effective couplings may remain phenomenologically viable while still producing significant modifications to neutron-star properties. Therefore, set-4 should be regarded primarily as an illustrative strong-coupling benchmark intended to demonstrate the possible astrophysical impact of a light vector mediator on the neutron-star equation of state, rather than as a fully realistic minimal dark-photon scenario.

For the nuclear matter EOS, we use the RMF model as in Eq. (\ref{eq:L_HAD_VEC}) with a parameterization considered recently in Ref. \cite{Kumar:2026tqe} obtained by using a Bayesian analysis. This parameter set for the RMF model is given in Table~\ref{tab:hadronic_matter_model_params} and corresponding nuclear matter saturation properties are given in Table~\ref{tab:hadronic_matter_saturation_properties}. 

\begin{table}[htb!]
    \centering
    \caption{Model parameters for DM particle $\chi$ and vector portal $Z'$ \cite{Taramati:2024kkn, Patel:2024zsu, Patra:2016ofq, Patra:2016shz}.}
    \begin{tabular}{lcccc}
    \hline \hline
     & $M_{\chi}$ & $M_{Z'}$ & $g_{NZ'}$ & $g_{\chi Z'}$ \\
    \hline
    set 1 & 200 GeV & 1800 GeV & 0.45 & 0.45 \\
    \hline
    set 2 & 1800 GeV & 900 GeV & 0.25 & 0.25 \\
    \hline
    {set 3} & {5 GeV} & {100 MeV} & {$5.0\times 10^{-4}$} & {0.80} \\
    \hline
    {\bf set 4} & {\bf 200 GeV} & {\bf 100\, MeV} & {\bf 0.45} & {\bf 0.45} \\
    \hline \hline
    \end{tabular}
    \label{tab:dark_matter_model_params}
\end{table}

\begin{table}[htbp]
    \centering
    \caption{The nucleon mass $(M)$ and the meson masses $m_{i}$ ($i=\sigma, \omega, \rho$) are taken as 939, 491.5, 782.5, and 763.0 MeV, respectively. The corresponding parameter set is adopted from Ref.~\cite{Kumar:2026tqe}, where it has been calibrated using the Bayesian analysis to reproduce the nuclear matter properties as well as NS observations. \label{tab:hadronic_matter_model_params}}
    \begin{tabular}{cc cc cc c}
    \hline \hline
    \multicolumn{1}{c}{$g_{\sigma N}$} & \multicolumn{1}{c}{$g_{\omega N}$} & \multicolumn{1}{c}{$g_{\rho N}$} & \multicolumn{1}{c}{$\kappa$} & \multicolumn{1}{c}{$\lambda$} & \multicolumn{1}{c}{$\xi_{\omega}$} & \multicolumn{1}{c}{$\Lambda_{\omega\rho}$} \\
    \hline
    8.8713 & 10.9532 & 9.3675 & 7.0597 & -0.0207 & 0.00083 & 0.0975 \\
    \hline \hline
    \end{tabular}%
\end{table}%

\begin{table}[htbp]
    \centering
    \caption{Nuclear saturation properties corresponding to the coupling constants listed in the previous table \ref{tab:hadronic_matter_model_params}. The tabulated quantities include the saturation density ($\rho_0$), binding energy per nucleon ($BE$), incompressibility ($K_0$), symmetry energy ($J_0$), and its slope ($L_0$) and curvature ($K_{\rm sym,0}$), evaluated consistently within the chosen parameter set. \label{tab:hadronic_matter_saturation_properties}}
    \begin{tabular}{cc cc cc cc cc}
    \hline \hline
    \multicolumn{1}{c}{$\rho_0\ ({\rm fm}^{-3})$} & \multicolumn{1}{c}{$\mbox{BE}$\ (MeV)} & \multicolumn{1}{c}{$K_0$ (MeV)} & \multicolumn{1}{c}{$J_0$\ (MeV)} & \multicolumn{1}{c}{$L_0$} & \multicolumn{1}{c}{$K_{\rm sym,0}$\ (MeV)} \\
    \hline
    0.148 & -15.757 & 250.933 & 24.345 & 39.414 & 52.684 \\
    \hline \hline
    \end{tabular}%
\end{table}%

\begin{figure}
    \centering
    \includegraphics[width=0.32\linewidth]{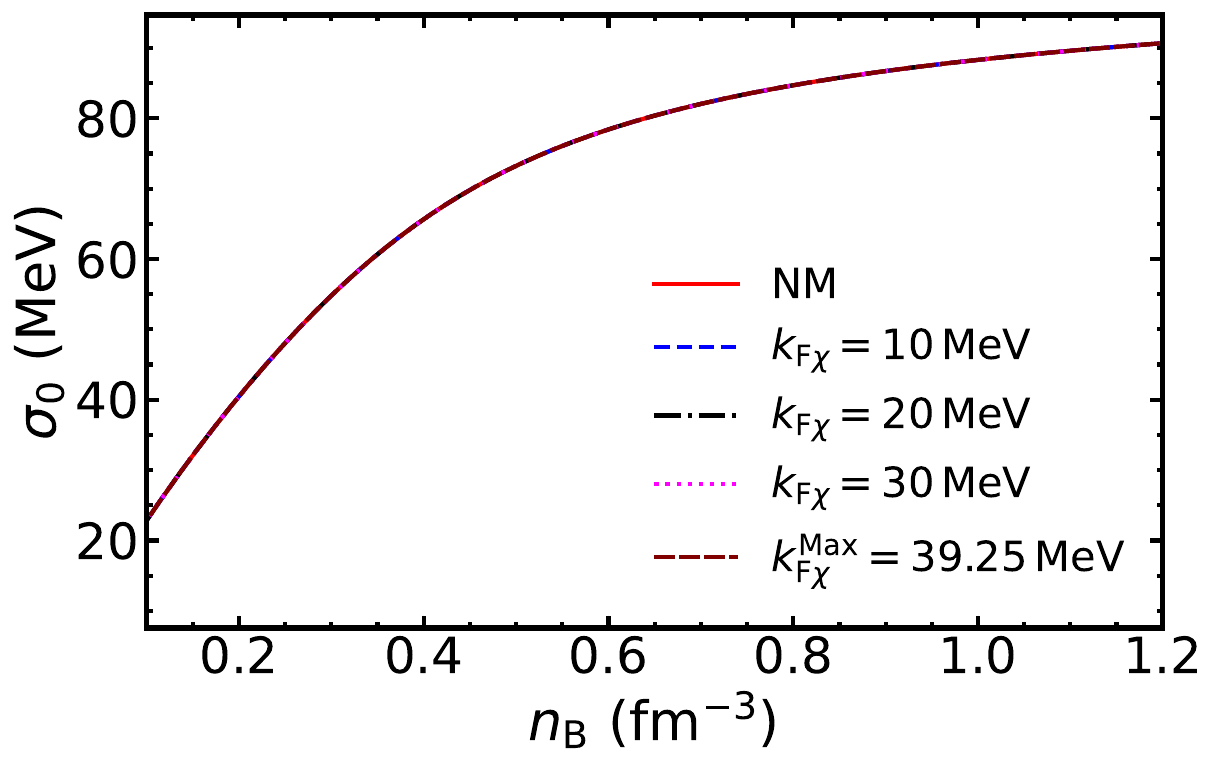}
    \includegraphics[width=0.32\linewidth]{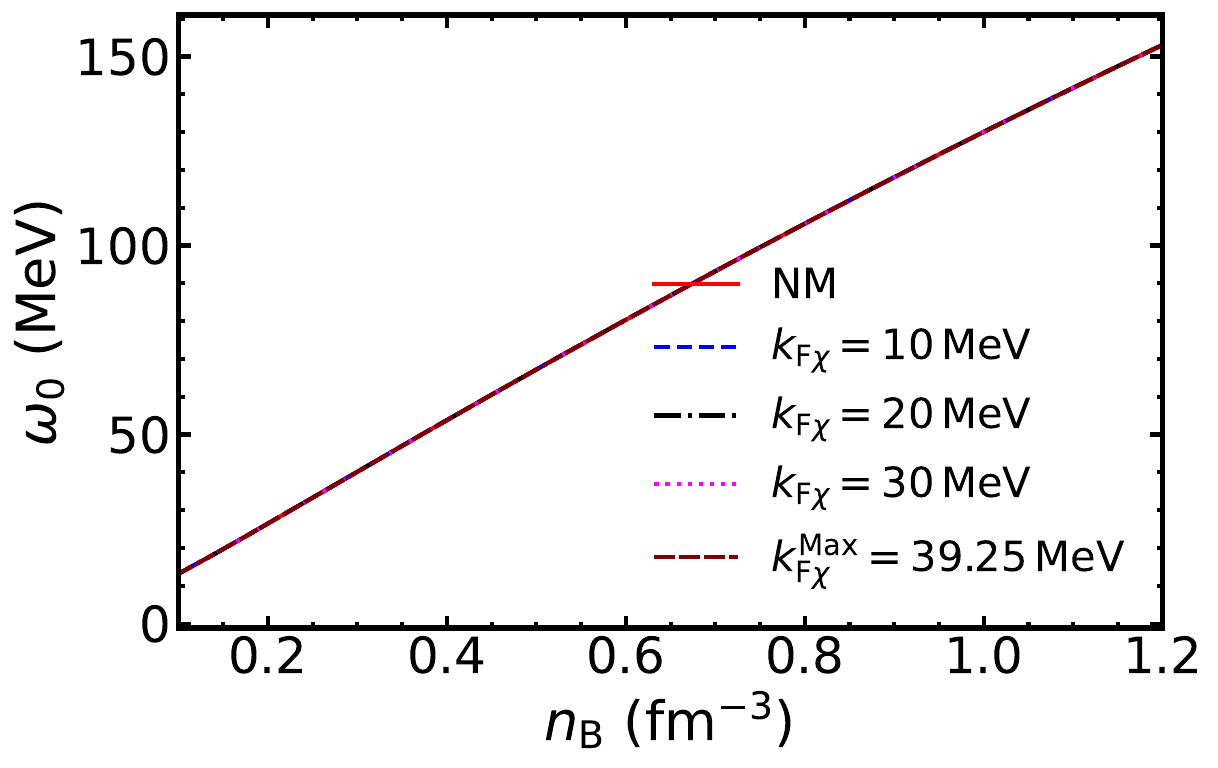}
    \includegraphics[width=0.32\linewidth]{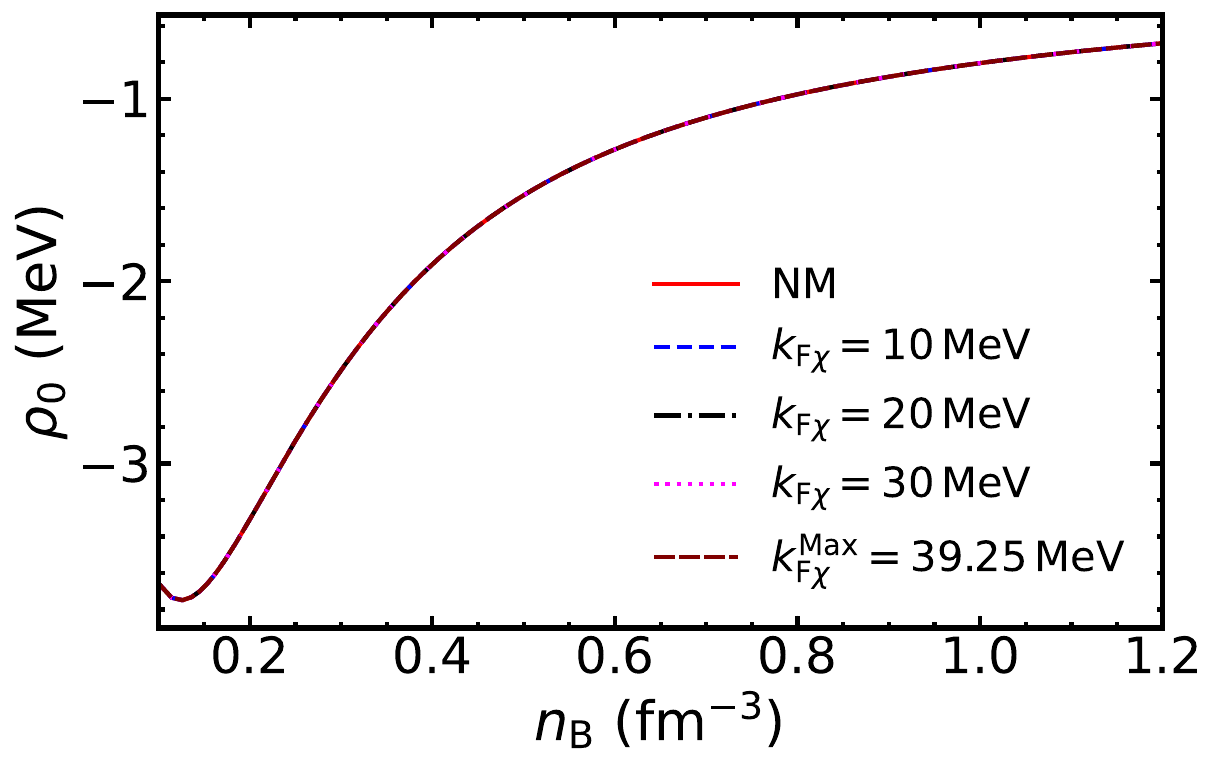}
    \caption{{Variation of the mean meson fields $\sigma_0$, $\omega_0$, $\rho_0$ as a function of baryon density $n_{B}$ for NS matter in the presence of vector portal DM for parameter set 1: (a) The left-panel corresponds to a scalar field $\sigma_0$, (b) middle-panel is for a vector meson field $\omega_0$, and (c) the right-penal corresponds to isovector field $\rho_0$ meson fields variations. The numerical results are shown for pure nuclear matter (NM) and NM with DM at different DM Fermi momenta $k_{\rm F\chi} = 10,\, 20,\ 30\, {\rm MeV}$ and $k_{F\chi}^{\rm Max}$.}}
    \label{fig:mfields}
\end{figure}
\begin{figure}
    \centering
    \includegraphics[width=0.32\linewidth]{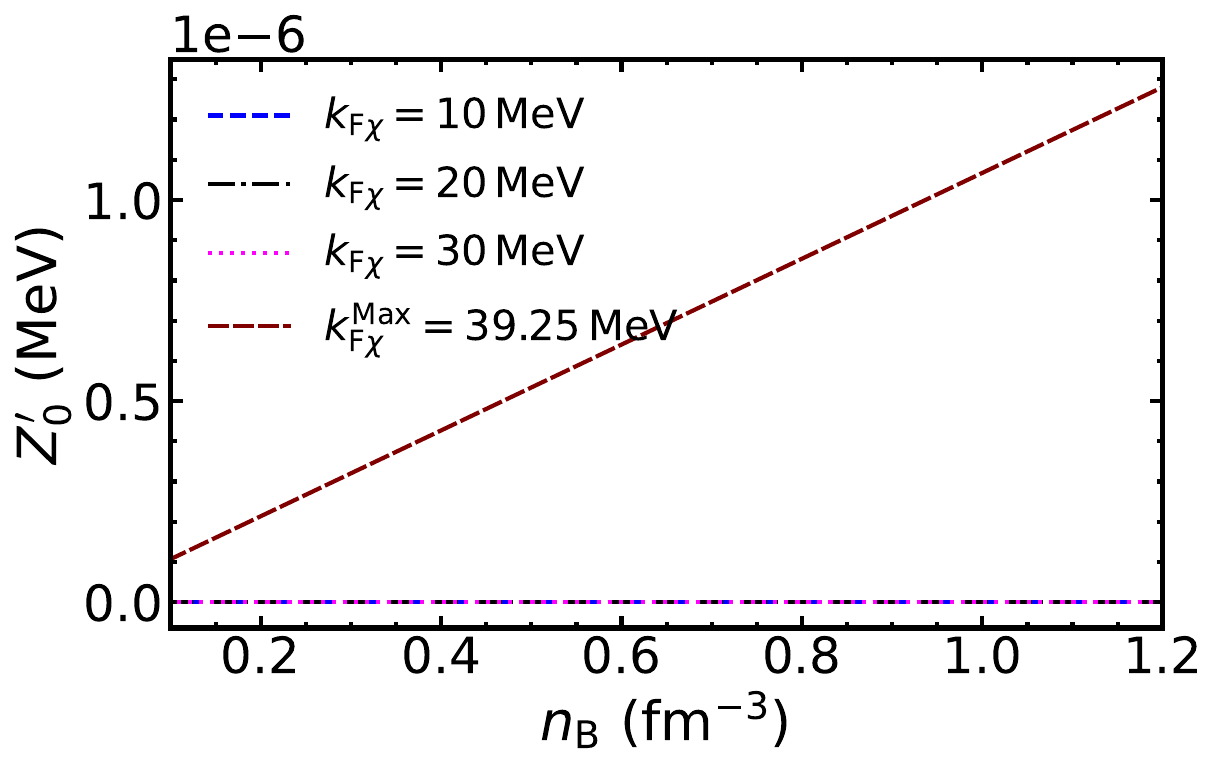}
    \includegraphics[width=0.32\linewidth]{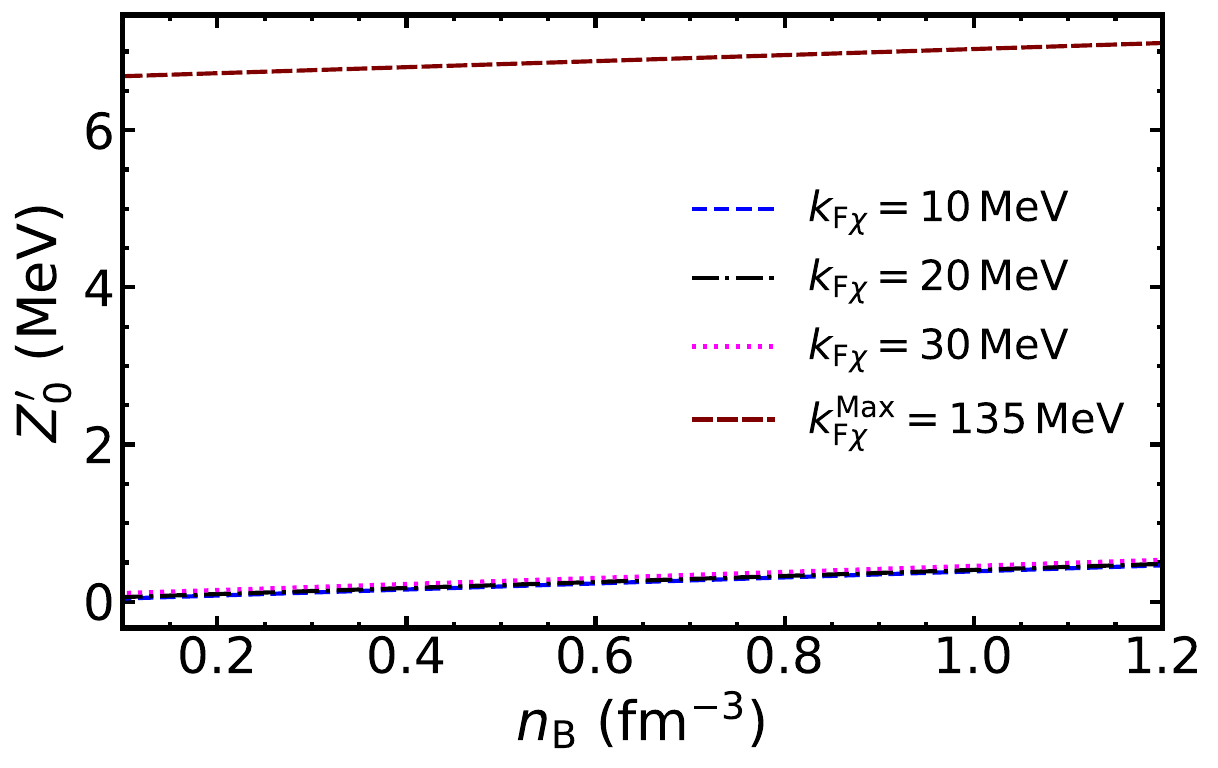}
    \includegraphics[width=0.32\linewidth]{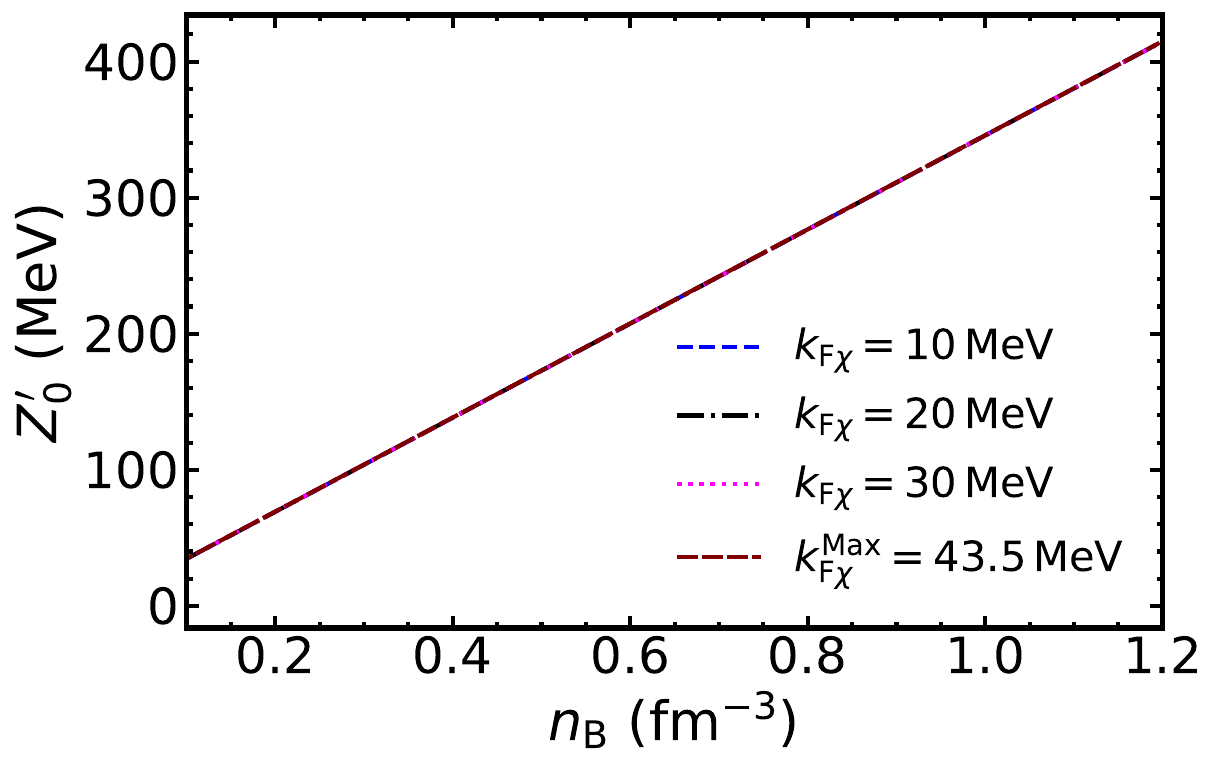}
    \caption{Mean-field value of the vector portal mediator $Z'_0$ as a function of baryon density $n_B$ for different DM Fermi momenta $k_{F\chi} = 10,\, 20,\, 30$ MeV and $k_{F\chi}^{\rm Max}$, shown for three representative parameter sets: (a) set 1 corresponding to a heavy mediator (left-panel), and (b) set 3 corresponding to a light mediator (middle-panel)with a smaller coupling $g_{\chi N}$, and (c) set 4 corresponding to a light mediator with larger coupling of the dark matter to ordinary matter}
    \label{fig:vfields}
\end{figure}
We next numerically estimate the mean fields for various mesons and vector boson using Eqs.(\ref{fieldeqns.sigma})-(\ref{fieldeqns.higgs}) for densities relevant for NSs. This is displayed in Fig. \ref{fig:mfields} for meson fields $\sigma_0$, $\omega_0$, $\rho_0$ as a function of baryon density $n_{B}$. Here, we have generated the plots using set 1 corresponding to a rather high vector boson mediator mass. {The behaviour for the set-2 parameters is also similar. We have plotted the meson mean fields. To discuss the impact of vector portal DM, we have also taken different values of Fermi momenta of DM i.e. $k_{\rm F\chi} = 10,\, 20,\,30\, {\rm MeV}$ which corresponds to different number densities of DM content in the NS matter. We have also plotted the same for $k_{F \chi}^{\rm Max}$ which correspond to maximum allowable value of the  DM fermi momenta constrained by astrophysical observations that we discuss later. As may be observed that the meson mean fields are rather insensitive to the DM parameters.  In Fig.\ref{fig:vfields}, we have plotted the mean field value for the vector boson $Z'_0$ as a function of baryon density $n_B$ for different sets of parameters. The left most panel corresponds to set-1 while the middle one and the right most one correspond the set 3 and set 4 respectively which correspond to a lighter vector boson mass $m_{Z'}=100\ \mbox{MeV}$.  It may be noted that for the densities, relevant for NSs, the vector boson mean fields turn out to be negligibly small for the heavy portal mass of set 1. This is also true for the case of set-2. Such a suppression is due to both a large value of the portal mass as well as the smaller coupling of the portal to the nucleon field. For lighter portal mass as in set-3 and set-4 cases, the mean portal field becomes comparatively significant. This is large particularly for set-4 which corresponds to a larger coupling of the portal to the nucleons. Thus, for set 1 and set 2, corresponding to a heavier $m_{Z'}$, the portal contribution to the energy density and the pressure becomes negligibly small. On the other hand, for set 3 and set-4,  the portal contribution to the energy density and pressure is relatively significant.}

\begin{figure}[h!]
    \centering
    \includegraphics[width=0.48\linewidth]{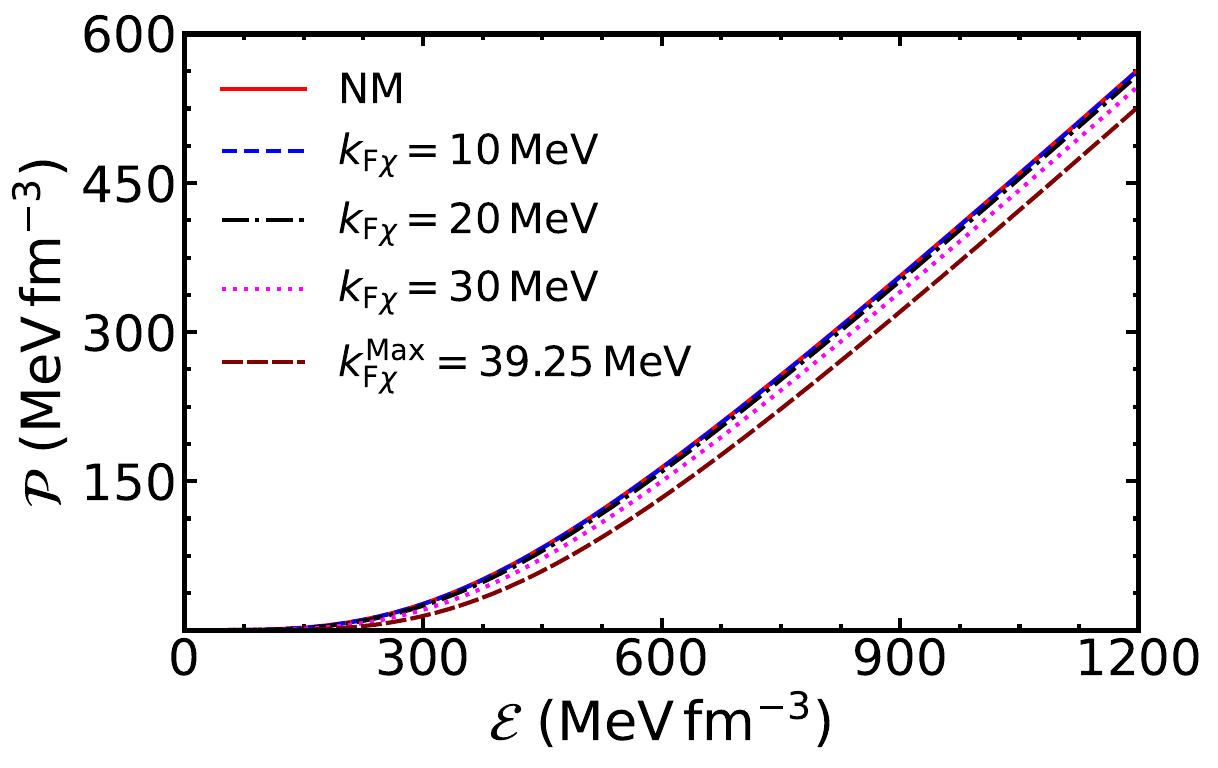}
    \includegraphics[width=0.48\linewidth]{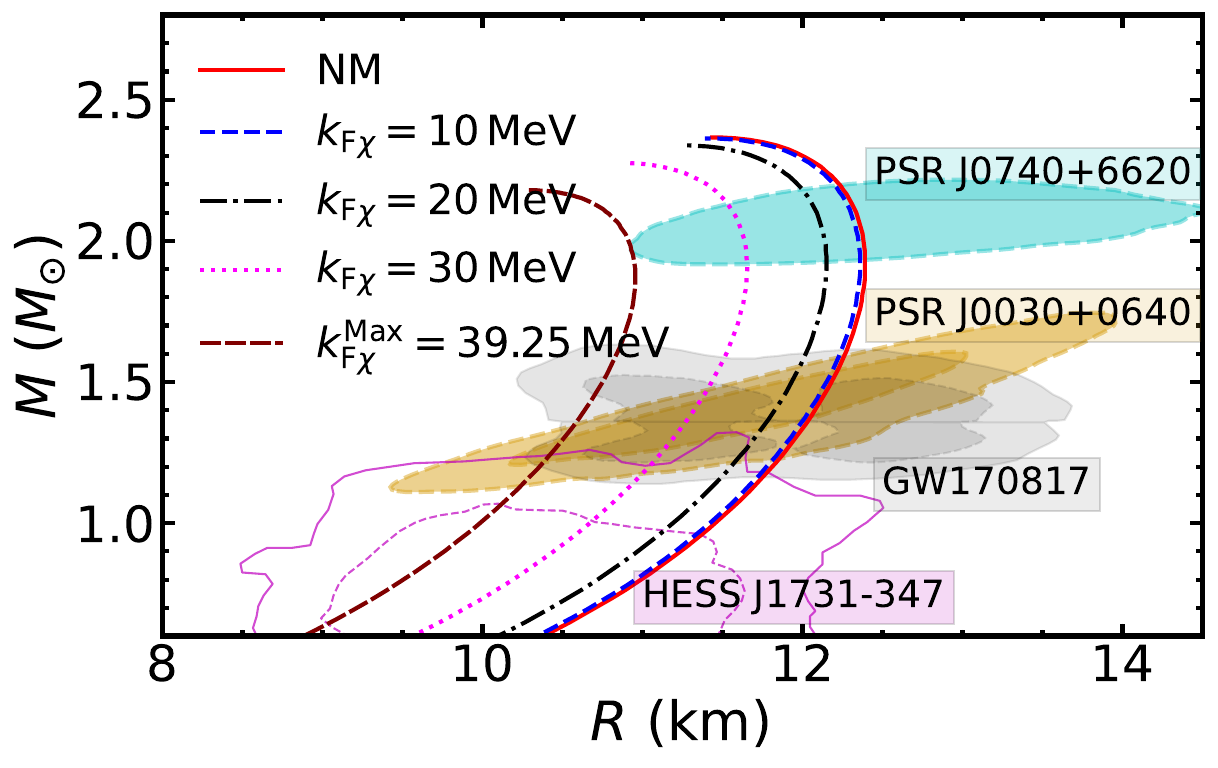}
    \caption{Equation of state (pressure $\mathcal{P}$ as a function of energy density $\mathcal{E}$) for NS matter admixed with fermionic DM via a vector portal interaction, shown for set 1 (left-panel). The corresponding mass radius relation ($M-R$ curve) is shown in the right-panel with various astrophysical observations (see text). 
     The red solid curve corresponds to pure nuclear matter (NM), while dashed, dot-dashed, and dotted curves represent DM admixture with Fermi momenta $k_{F\chi} = 10,\, 20,\, 30$ MeV, respectively.}
    \label{fig:EoS-mr-set1}
\end{figure}

Before proceeding further, it is important to discuss the validity of employing the RMF parameter set listed in Table~\ref{tab:hadronic_matter_model_params}, which was originally determined from nuclear saturation properties and observational constraints on neutron stars. Ideally, in the presence of dark matter admixed matter, the RMF parameters should be refitted by incorporating the additional vector portal contribution into the determination of the saturation properties. Such a self-consistent analysis, however, is beyond the scope of the present work. Nevertheless, we provide a posterior justification for the use of the same RMF parameter set in the dark-matter-admixed neutron star matter considered here.

It is worth noting that the saturation properties of nuclear matter are primarily governed by the effective nucleon mass, $M^{*}$, and the effective chemical potential, $\mu^{*}$. To leading order, the effective nucleon mass does not receive any direct contribution from the vector portal interaction. The dominant portal effect enters through the effective chemical potential, which depends linearly on the vector portal contribution, as shown in Eq.~(\ref{effective-chemical-potential-nl3}). For benchmark parameter sets~1 and 2, the mean-field value of the portal mediator, $Z'_0$, is negligibly small as may be observed in left most panel of Fig.\ref{fig:vfields}. Consequently, the vector portal contribution to the effective chemical potential is insignificant, and the nuclear saturation properties remain essentially unchanged. Therefore, a refit of the RMF parameters may not be required in these cases.

The set~3 and set~4 benchmark parameters correspond to a light vector portal. For set-3, the coupling of the mediator to the visible sector is highly suppressed ($g_{qZ'}\sim10^{-4}$), while the dark-sector coupling remains sizable ($g_{\chi Z'}\simeq0.8$) \cite{Taramati:2024kkn, Patel:2024zsu, Patra:2016ofq, Patra:2016shz}. As a result, the corresponding portal mean field remains relatively small, of order $\mathcal{O}(\mathrm{MeV})$, and does not significantly alter the saturation properties of nuclear matter. Therefore, the original RMF parameterization can still be used consistently in this case.

On the other hand, for sufficiently large dark matter densities, corresponding to larger values of $k_{F\chi}$, the portal mean field, $Z'_{0} \propto g_{\chi Z'}\,n_\chi$, can become sizeable and may eventually influence the saturation properties of nuclear matter. In such a regime, a complete refit of the RMF parameters would in principle be required for both set~3 and set~4. For set~4, the vector portal contribution near saturation density is approximately $g_{NZ'}Z'_{0} \simeq 0.45 \times 60~{\rm MeV} \simeq 27~{\rm MeV}$, as inferred from the right most panel of Fig.~\ref{fig:vfields}, whereas the conventional RMF contribution is $g_{\omega N} \omega_0 \simeq 400~{\rm MeV}$. Thus, the portal contribution remains more than an order of magnitude smaller than the dominant RMF vector interaction near saturation density. This justifies, to a good approximation, the use of the same RMF parameterization in the present exploratory study.

At higher baryon densities relevant to neutron-star interiors, however, the portal contribution can become sufficiently large to modify the equation of state and consequently affect neutron-star observables such as the mass-radius relation and tidal deformability. While a complete refit of the RMF parameters would be desirable, particularly for the set-4 benchmark, for a fully self-consistent treatment of this benchmark, the dominant features of the nuclear saturation properties remain largely unaffected. We therefore employ the same RMF parametrization as a reasonable approximation in the present investigation.

\begin{figure}[t!]
    \centering
    \includegraphics[width=0.48\linewidth]{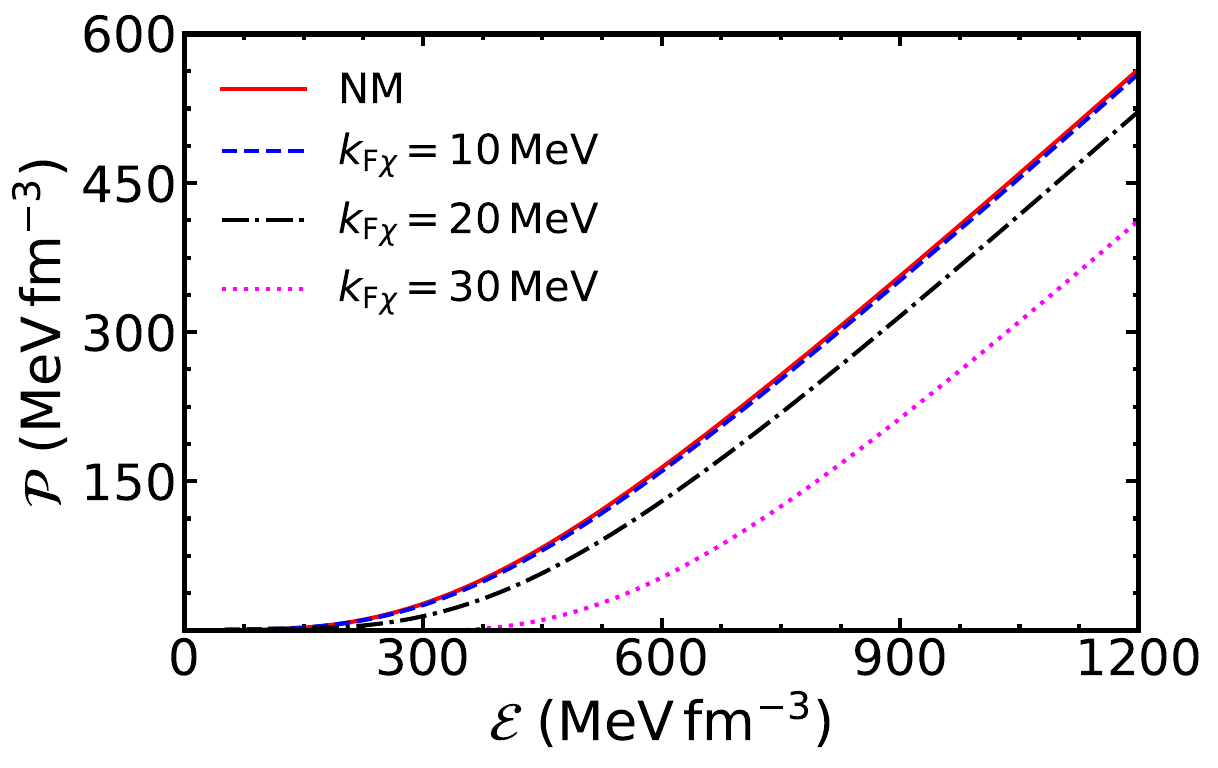}
    \includegraphics[width=0.48\linewidth]{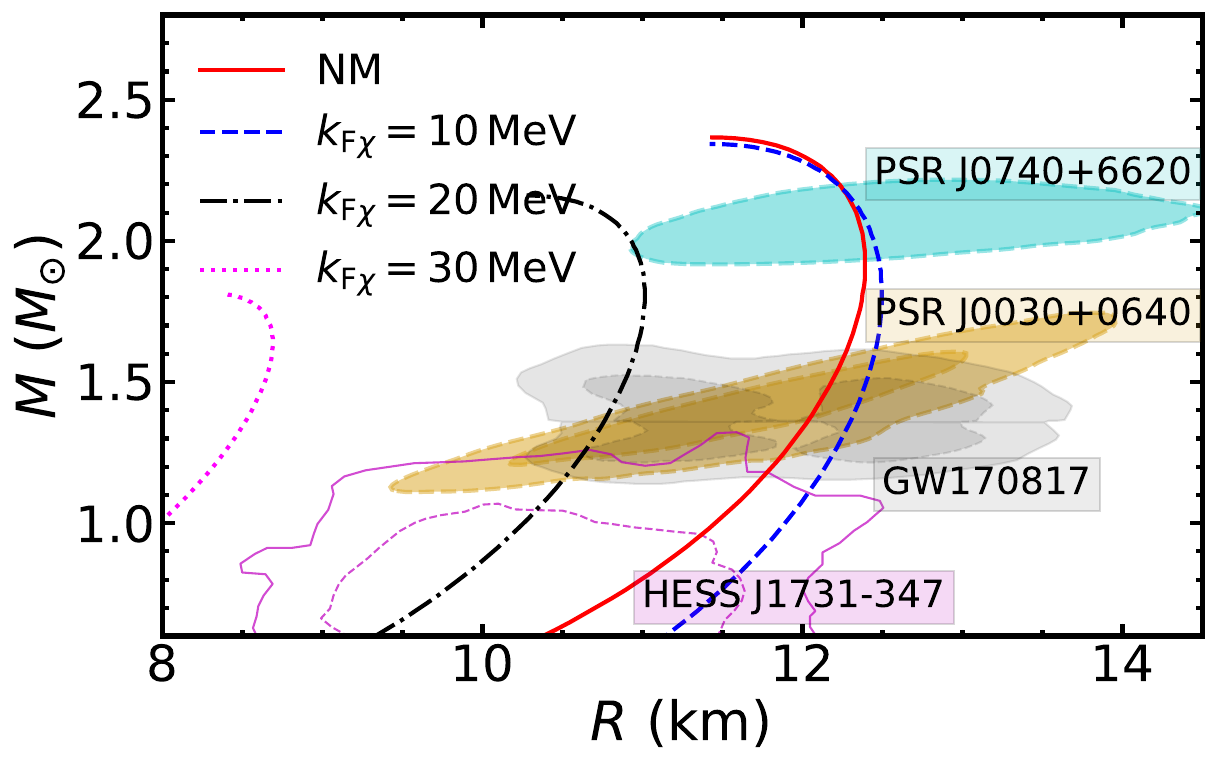}
    \caption{EoS for NS matter admixed with fermionic DM via a vector portal interaction, shown for set 2 (left-panel). The corresponding $M-R$ curve is shown in the right-panel. The red solid curve corresponds to pure nuclear matter (NM), while dashed, dot-dashed, and dotted curves represent DM admixture with Fermi momenta $k_{F\chi} = 10,\, 20,\, 30$ MeV, respectively.}
    \label{fig:EoS-mr-set2}
\end{figure}

\begin{figure}[htb!]
    \centering
    \includegraphics[width=0.48\linewidth]{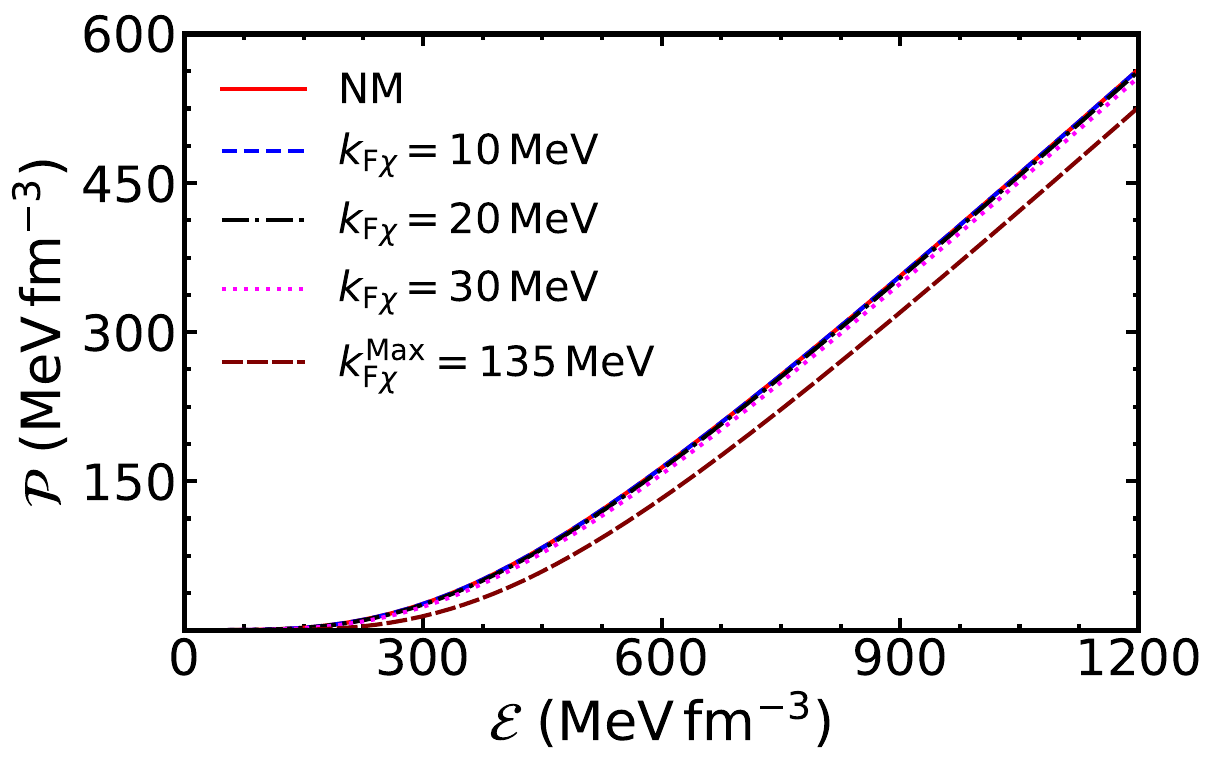}
    \includegraphics[width=0.48\linewidth]{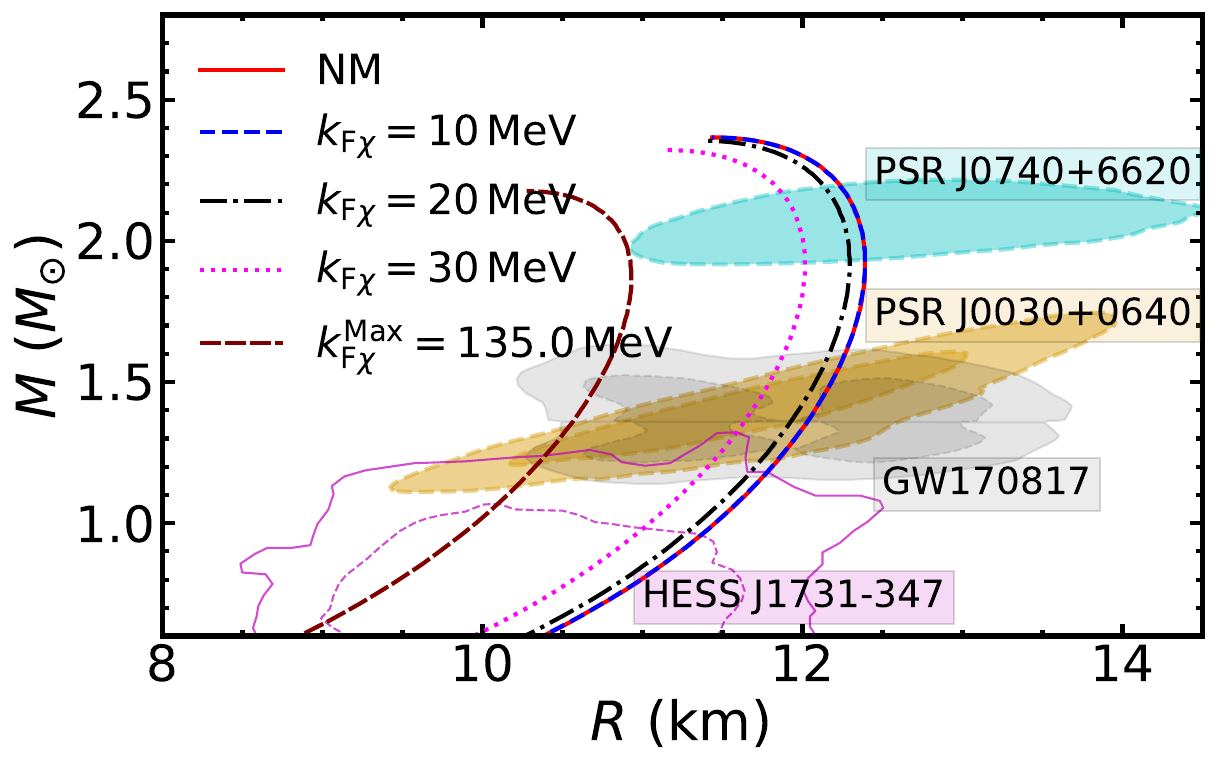}
    \caption{Equation of state (left panel) and corresponding mass–radius relations (right panel) for NS matter admixed with fermionic DM via a vector portal interaction for set 3, corresponding to a light mediator ($M_{Z^\prime} = \mbox{100\,MeV}$) and couplings are as $g_{\chi Z^\prime} = 0.80$ and $g_{N Z^\prime} = 5.0\times 10^{-4}$. The red solid curves represent pure nuclear matter (NM), while dashed, dot-dashed, and dotted curves correspond to DM admixture with Fermi momenta $k_{F\chi} = 10,\, 20,\, 30$ MeV respectively. Brown long-dashed curve corresponds to the maximum DM Fermi-momentum $k_{F\chi}^{\rm Max} = 135$ MeV.}
    \label{fig:EoS-mr-set3}
\end{figure}

\begin{figure}[htb!]
    \centering
    \includegraphics[width=0.48\linewidth]{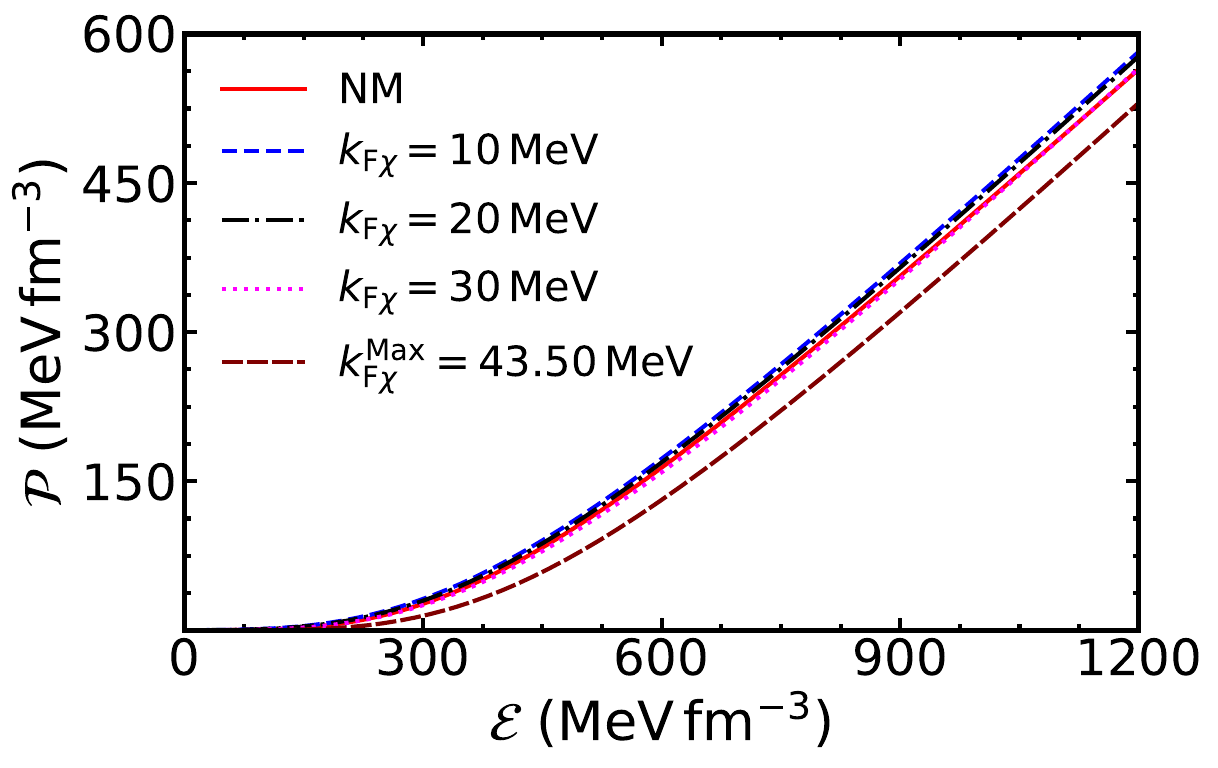}
    \includegraphics[width=0.48\linewidth]{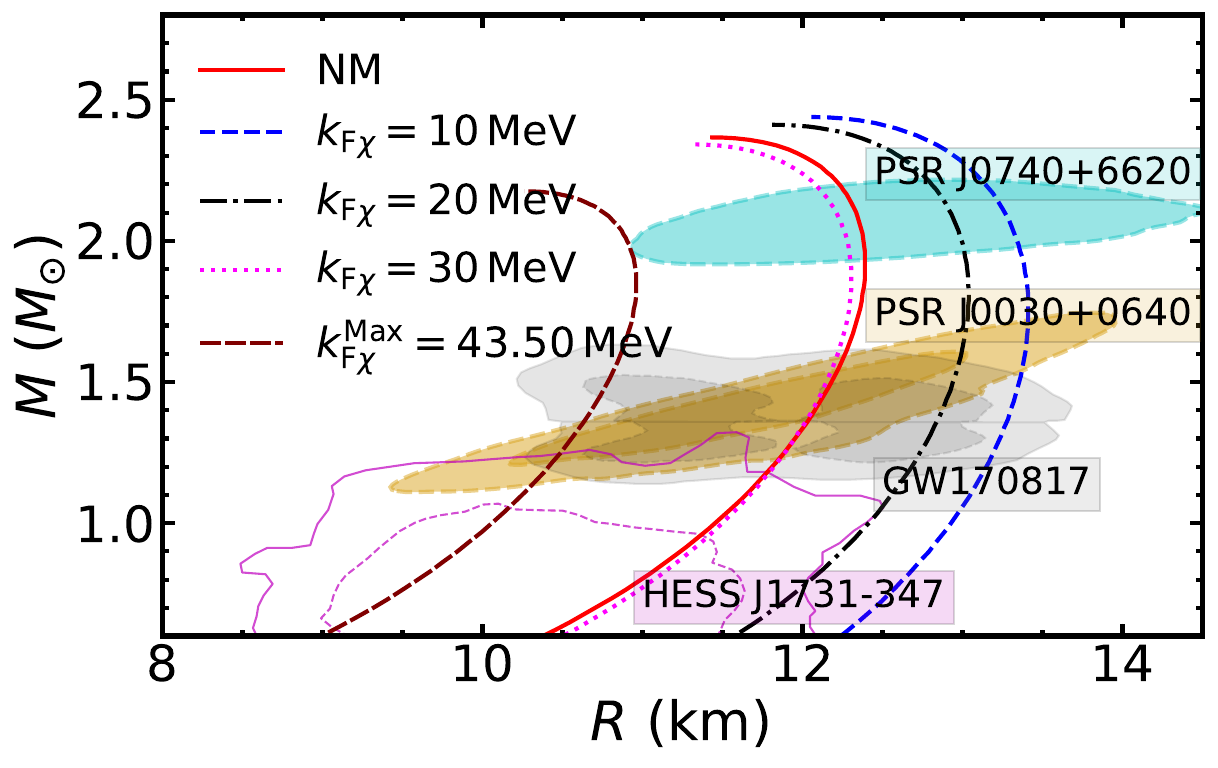}
    \caption{Equation of state (left panel) and corresponding mass–radius relations (right panel) for NS matter admixed with fermionic DM via a vector portal interaction for set 4, corresponding to a light mediator ($M_{Z^\prime} = \mbox{100\,MeV}$). The red solid curves represent pure nuclear matter (NM), while dashed, dot-dashed, and dotted curves correspond to DM admixture with Fermi momenta $k_{F\chi} = 10,\, 20,\, 30$ MeV respectively.}
    \label{fig:EoS-mr-set4}
\end{figure}

With these comments regarding the RMF model, we next discuss the EOS and the mass-radius relations for each of the four scenarios. Using the parameter sets (for DM matter in Table \ref{tab:dark_matter_model_params} and for hadronic matter in Table \ref{tab:hadronic_matter_model_params}), we calculate the EOS as defined in Eqs. (\ref{energy_density_nm}) and (\ref{pressure_nm}). Once the EOS is obtained, we numerically solve the TOV Eqs. (\ref{tov_pressure}) and  (\ref{tov_mass}) along with the tidal deformability relation as given in Eq. (\ref{tidal_y}) simultaneously to obtain the mass-radius as well as tidal deformability-mass relations. In Figs. \ref{fig:EoS-mr-set1}-\ref{fig:EoS-mr-set4}, we present the EoSs (left-panel) and the corresponding mass radius relation (right-panel) for all the four sets of DM parameters. The results are presented alongside the purely hadronic NSs corresponding to 'without DM' case represented by the red solid line. Let us first discuss the behaviour of the  EOSs for each of the four  sets. As noted earlier, the mean field for the vector boson  has negligible contribution to the pressure and  energy density for heavier $Z^\prime$ mass. Thus, for such cases, the effect of DM on the EoSs in Figs. \ref{fig:EoS-mr-set1}-\ref{fig:EoS-mr-set2} is primarily determined by the DM matter mass ($M_{\chi} = 200\,\mbox{GeV},\ 1800\ {\rm GeV}$) and their densities i.e.  the values of their corresponding fermi momenta $k_{F\chi}$. As may be noted from the EoSs , increasing the DM Fermi-momentum softens the EOS i.e. a reduction in pressure support and shifting the EOS to higher energy densities in all the plots. Further, the impact of $k_{\rm F\chi}$ is more pronounced for higher DM mass. We may note here that this analysis considers only NS core and neglects the crust contributions.

The EOS for set 3 with hierarchical couplings of the portal, i.e., $g_{\chi Z'}>>g_{NZ'}$, is shown in Fig.\ref{fig:EoS-mr-set3}. Here, the portal contribution is small but non-negligible. Reducing the portal mass hardens the EOS. The EOS for set 3 becomes harder compared to set 1 and set 2, but it is still softer compared to the `without DM' case. However, with higher $k_{F\chi}$, the EOS becomes softer. In Fig. \ref{fig:EoS-mr-set4}, we display the effect of a light vector portal mass and a stronger coupling of the portal to nucleons, i.e., for set-4, on the EOS. Here,  for smaller densities, i.e., smaller $k_{F\chi}$, the EOS becomes stiffer compared to the 'without DM' case, and with higher $k_{F\chi}$, the EOS becomes softer. In Fig.\ref{fig:EoS-mr-set4}, on the left panel, the EOS is displayed, which corresponds to a stronger coupling of the portal field to the nucleons, which was motivated by the asymmetric DM scenario. Here, for smaller values of DM densities, the EOS can become stiffer even from the 'without DM' scenario, which can be seen, e.g., for $k_{F\chi}=10$MeV. Here, the contribution to the pressure from the portal is non-negligible; the equation of state becomes stiffer at higher densities than in the case of a heavier mediator mass. This is clearly demonstrated in the appendix for the same case of a Dirac fermion with a vector interaction only. This arises essentially because, at lower densities, the pressure scales as $\sim n_B^{4/3}$, while at higher densities it scales quadratically (${\mathcal{P}}\sim n_B^2$) with the net fermion density. With increasing $k_{F\chi}$, the EOS becomes softer as is seen Fig.\ref{fig:EoS-mr-set4}.

\begin{table}[htb!]
    \centering
    \caption{{Stellar properties for four distinct parameter sets as given in Table \ref{tab:dark_matter_model_params}, each evaluated at four DM Fermi momenta ($k_{{\rm F}\chi} = 10,\ 20,\ 30\ {\rm MeV}$ and $k_{F\chi}^{\rm Max}$). For every configuration, we report the maximum mass, the corresponding radius, and the tidal deformability for a $1.4\ M_{\odot}$ neutron star. We also report the same for pure nuclear matter case. \label{tab:staller_properties}}}
    \begin{tabular}{|c|cccc|}
    \hline\hline
     & \multicolumn{1}{l}{$k_{{\rm F}\chi}\ ({\rm MeV})$} & \multicolumn{1}{l}{$M_{\rm Max}\ ({M_{\odot}})$} & \multicolumn{1}{l}{$R_{\rm Max}\ ({\rm km})$} & \multicolumn{1}{l|}{$\tilde{\Lambda}_{1.4}$} \\
     \hline
     \multirow{1}[0]{*}{Nuclear matter} 
     & --- & 2.37 & 11.44 & 393.40 \\
     \hline
    \multirow{3}[0]{*}{set 1} 
     & 10 & 2.36 & 11.39 & 388.97 \\
     & 20 & 2.34 & 11.27 & 360.44 \\
     & 30 & 2.28 & 10.89 & 298.41 \\
     & 39.25 & 2.18 & 10.29 & 224.3 \\
     \hline
    \multirow{3}[0]{*}{set 2} 
     & 10 & 2.34 & 11.42 & 386.19 \\
     & 20 & 2.16 & 10.28 & 221.98 \\
     & 30 & 1.81 & 8.38  & 73.89 \\
     \hline
     \multirow{4}[0]{*}{set 3} 
     & 10 & 2.44 & 12.06 & 531.56 \\
     & 20 & 2.41 & 11.81 & 480.21 \\
     & 30 & 2.34 & 11.33 & 378.32 \\
     & 135.0 & 2.18 & 11.28 & 222.21 \\
     \hline
    \multirow{3}[0]{*}{set 4} 
     & 10 & 2.44 & 12.06 & 531.56 \\
     & 20 & 2.41 & 11.81 & 480.21 \\
     & 30 & 2.34 & 11.33 & 378.32 \\
     & 43.5 & 2.18 & 10.26 & 224.16 \\
     \hline
     \hline
    \end{tabular}%
    \label{tab:nsprops}%
\end{table}%

We next discuss the mass-radius relations for vector-portal DM admixed NS matter for the four cases given in Table \ref{tab:dark_matter_model_params}. The corresponding stellar properties are summarized in Table~\ref{tab:staller_properties} for all the four cases. For comparison, we also show the results for the case of pure nuclear matter in Table~\ref{tab:staller_properties}. We have plotted the same (M-R) in the right panels of Figs. \ref{fig:EoS-mr-set1}-\ref{fig:EoS-mr-set3}. In the same figures, we also display different observational results.  For the largest NS mass observed till now i.e. with mass $2.08 \pm 0.07\ {M}_{\odot}$~\cite{Dittmann:2024mbo} at 68\% confidence interval for the compact star, PSR J0740+6620, is shown as the cyan band with dotted outline. We also display the bayesian parameter estimation of the mass and equatorial radius of the millisecond pulsar PSR J0030+0451 as reported by the NICER mission~\cite{Vinciguerra:2023qxq} shown as the yellow regions. Apart from the NICER data, we also display the constraints from data extracted from the gravitational wave observations (GW170817) in LIGO/Virgo~\cite{LIGOScientific:2017vwq, LIGOScientific:2017ync} in the gray color. The outer (light gray) and inner (dark gray) regions indicate the 90\% (solid) and 50\% (dashed) confidence intervals of LIGO/Virgo analysis for each binary component of GW170817 event~\cite{LIGOScientific:2018cki}. Along with these observations, we also display the lightest known compact stars recently observed in HESS J1731-347~\cite{Doroshenko:2022nwp} observation with a mass and radius measurement $0.77^{+0.20}_{-0.17}\ {\rm M}_{\odot}$ and $10.4^{+0.86}_{-0.78} {\rm km}$, respectively, by the pink dashed contour lines.

Let us first discuss the M-R relation for the cases when the portal mass is large, i.e., the results for set 1 and set 2, which are displayed 
in the right panels of Figs \ref{fig:EoS-mr-set1}-\ref{fig:EoS-mr-set2}. Both figures illustrate that increasing either the Fermi momentum $k_{\rm F\chi}$ or the mass $M_{\chi}$ of DM reduces both the maximum mass and radius of NSs. As displayed in Fig.\ref{fig:EoS-mr-set1}, it may be observed that for $M_{\chi} = 200\, {\rm GeV}$ (set 1) the results for $k_{F\chi}$=10, 20 and 30 MeV, the mass and radius appear to satisfy all observational constraints. As shown in Fig.~\ref {fig:EoS-mr-set1}, increasing the fermi momentum of DM shifts the M-R curve to the left, and eventually, for sufficiently large $k_{F\chi}$, it fails to meet the constraints from NICER PSRJ0740+6620. As mentioned earlier, increasing the DM contributions (i.e., increasing $k_{F\chi}$) reduces the pressure support, making the EOS softer, leading to a lower maximum mass. For the higher DM mass i.e.  for $M_{\chi} = 1800\, {\rm GeV}$ (set 2), as may be seen in Fig.\ref{fig:EoS-mr-set2}, while the lower $k_{F\chi}$ results satisfy all the existing observational constraints, the higher Fermi momentum ($k_{\rm F\chi} =30\, {\rm MeV}$) of DM fails to satisfy any of the observational constraints. {In Fig.\ref{fig:EoS-mr-set3}, we show the results for the case of light vector mediator mass, i.e., $m_{Z'} \sim 100$~MeV, and hierarchical couplings to DM and NM corresponding to set 3 parameters. As mentioned earlier, in this case, the portal's contribution to the EOS becomes non-negligible, making the EOS comparatively stiffer than with a larger portal mass, as in sets 1 and 2. The EOS, however, is softer compared to the 'without DM' case. This leads to a smaller maximum mass and a smaller radius as the dark matter density increases. For such a set of parameters, the maximum value of $k_{F\chi}$ 
turns out to be $k_{F\chi}=135$~MeV that satisfies NICER and GW170817 observations. Thus, compared to sets 1 and 2, a smaller mediator mass can accommodate higher DM densities inside NSs, consistent with all observational constraints. Fig.\ref{fig:EoS-mr-set4} shows the mass-radius relations for set 4. In this case, both the light portal mass and a stronger $g_{NZ'}$ make the EOS stiffer as compared to the "without DM" case for smaller DM fermi momenta. 
This leads to a larger mass and radius for DM admixed NSs as compared to purely hadronic stars for $k_{F\chi}=10$ and $20$~MeV. On the other hand, the introduction of DM leads to softening of the EoS. Thus, for a larger Fermi momentum of DM, i.e., $k_{\rm F\chi} =30\, {\rm MeV}$, the mass and radius get reduced. The maximum DM density that can be consistent with observational constraints turns out to be with corresponding Fermi momentum $k_{F\chi}=43.5$~MeV.}

\begin{figure}[t!]
    \centering
    \includegraphics[width=0.48\linewidth]{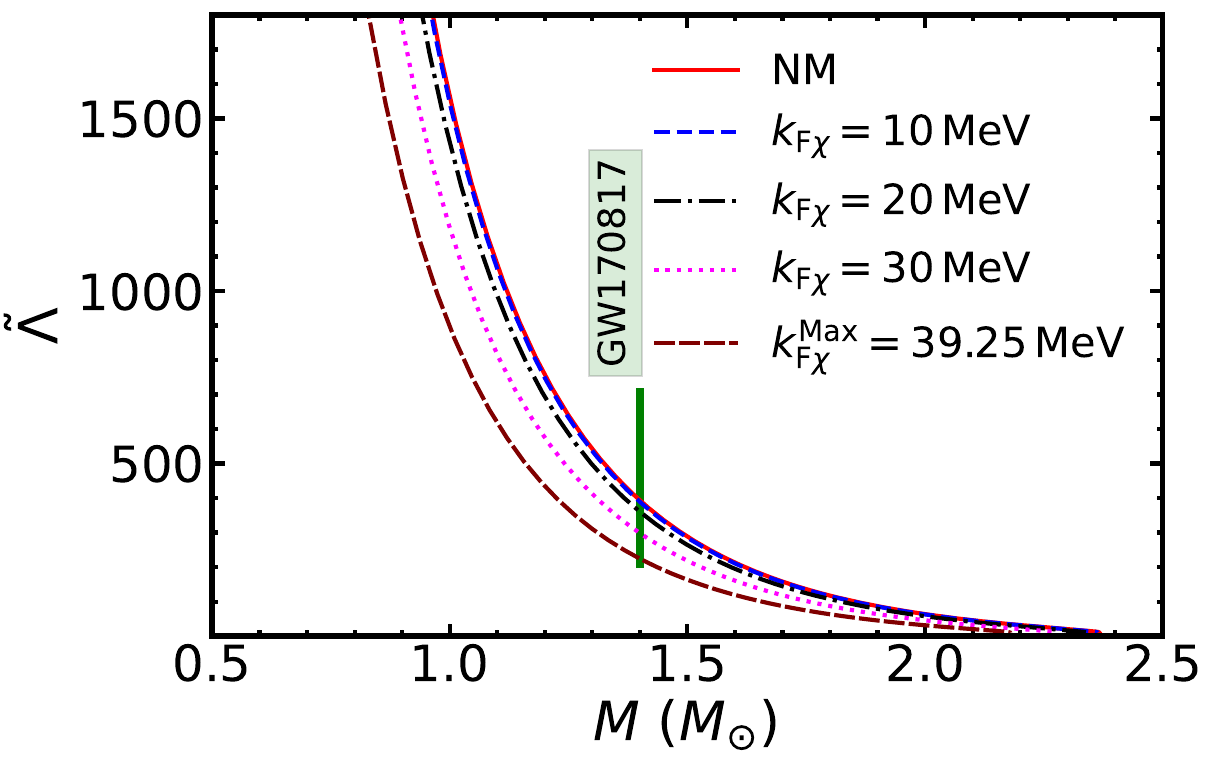}
    \includegraphics[width=0.48\linewidth]{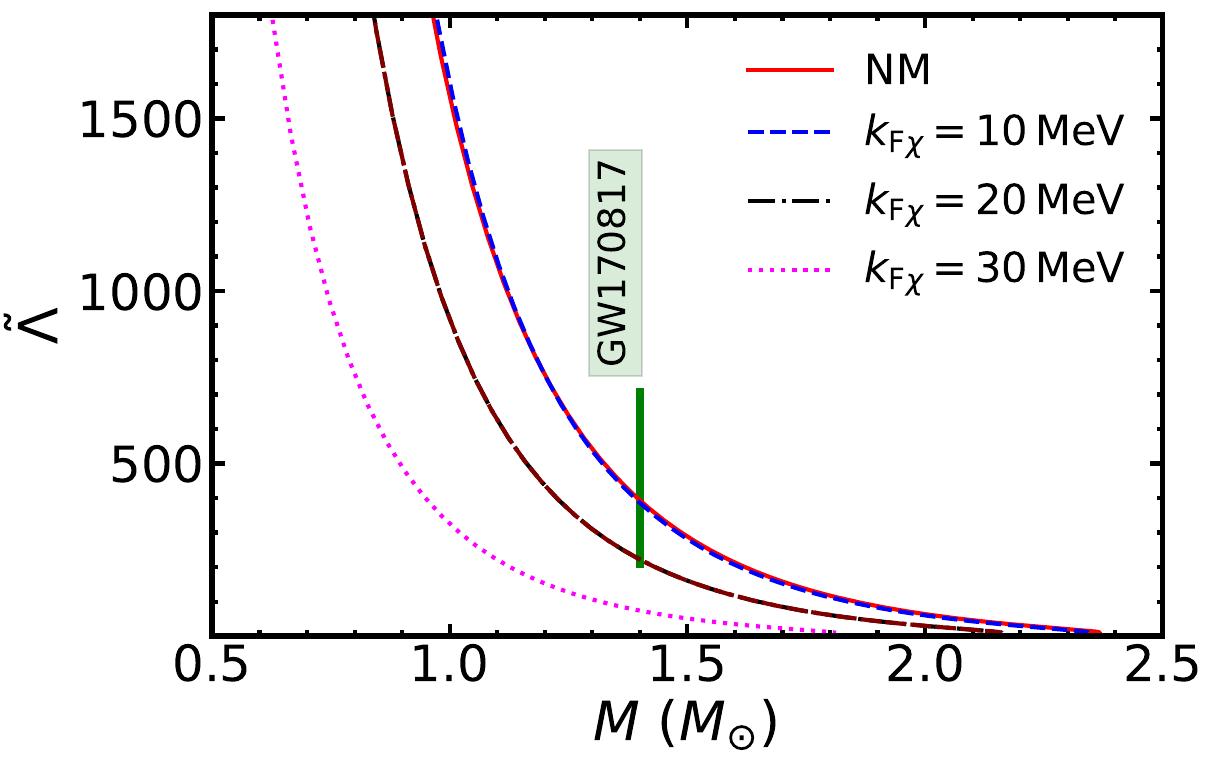}
    \caption{Dimensionless tidal deformability $\Lambda$ as a function of NS mass $M$ for set 1 (left panel) and set 2 (right panel) parameter choices in the presence of vector portal DM. The red solid curve corresponds to pure nuclear matter (NM), while dashed, dot-dashed, and dotted curves represent DM admixture with Fermi momenta $k_{F\chi} = 10,\, 20,\, 30$ MeV and $k_{F\chi}^{\rm Max}$, respectively.}
    \label{fig:atidal_set1_and_set2}
\end{figure}
\begin{figure}[t!]
    \centering
    \includegraphics[width=0.48\linewidth]{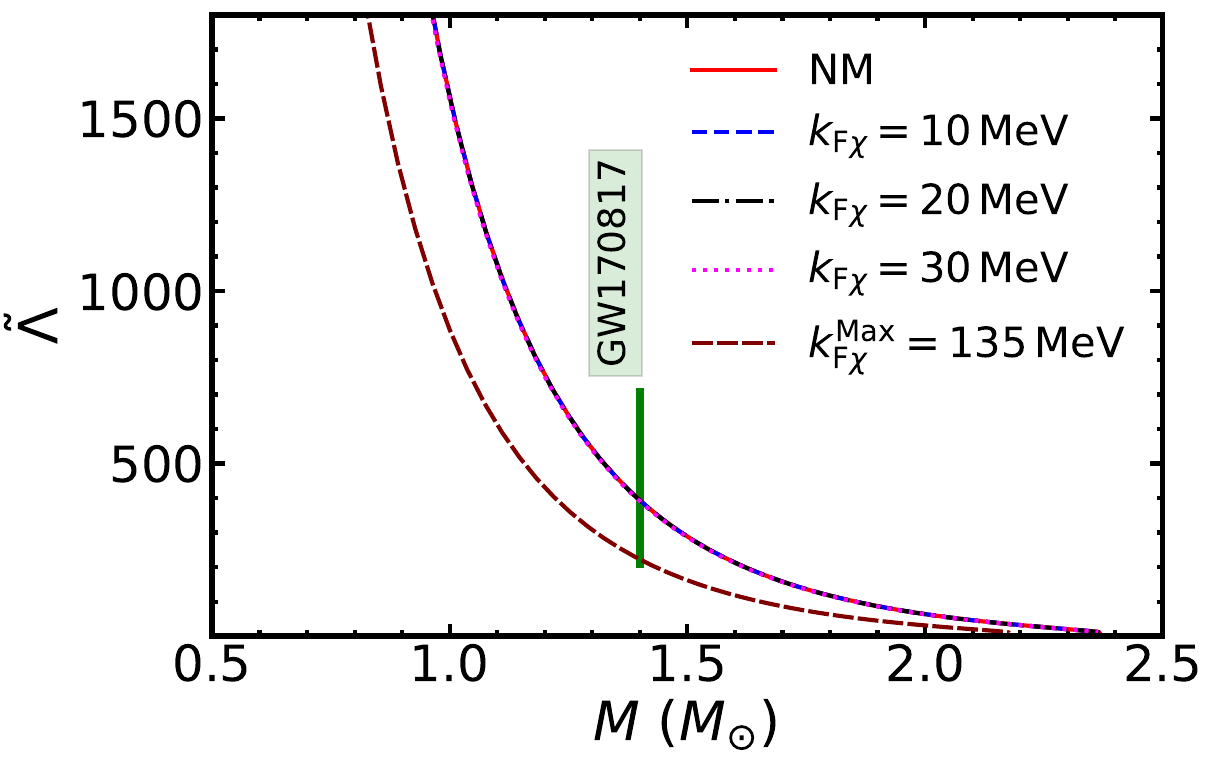}
    \includegraphics[width=0.48\linewidth]{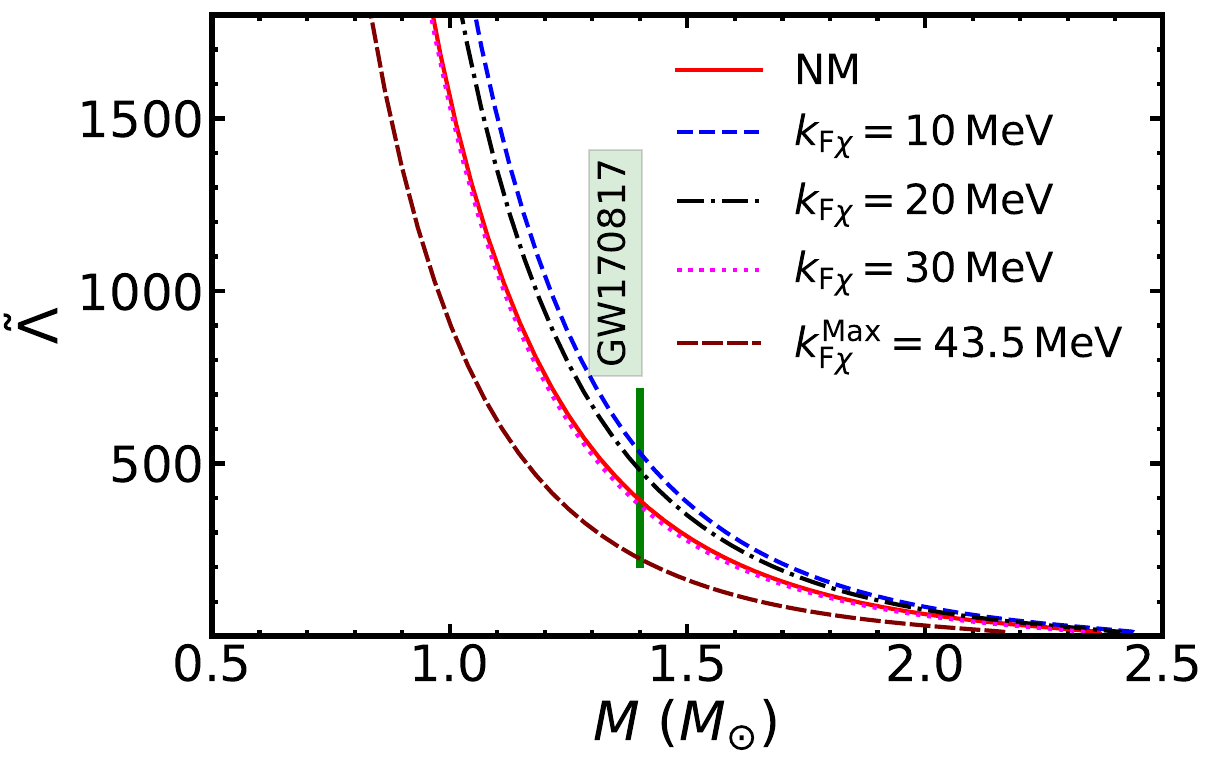}
    \caption{Dimensionless tidal deformability ($\tilde{\Lambda}$) as a function of NS mass for set 3 (left) and set 4 (right) at Fermi momentum ($k_{{\rm F}\chi} = 10,\,20,\,30\ \mathrm{MeV}, k_{F\chi}^{\rm Max}$) and the maximum possible Fermi momentum $k_{F\chi}^{\rm Max}$.}
    \label{fig:atidal_set3}
\end{figure}

Using Eqs.(\ref{love_number_k2})-(\ref{eq:tidal-final}), we next intend to discuss how the tidal deformability of a NS manifests important information about its internal structure and the equation of state (EoS). The numerical results for the relation between the tidal deformability and mass of the star are presented in Fig.s~\ref{fig:atidal_set1_and_set2} corresponding to a heavier portal mass and Fig.s~\ref{fig:atidal_set3}. We also show here the constraints on the tidal deformability of an NS with mass 1.4 M$_\odot$ from GW170817 ($\tilde\Lambda=190^{+390}_{-120}$~\cite{LIGOScientific:2017vwq, LIGOScientific:2017ync}). Fig.\ref{fig:atidal_set1_and_set2} displays the results for a heavy vector mediator mass corresponding to set 1 (left panel) and set 2 (right-panel). With increasing DM fractions, i.e. $k_{F\chi}$, the compactness parameter increases and makes the star less sensitive to tidal forces. This results in a decrease of tidal deformability parameter as may be seen in left-panel of Fig.\ref{fig:atidal_set1_and_set2}. At higher DM mass ($M_\chi$) shown in right-panel of Fig.\ref{fig:atidal_set1_and_set2}. {Here, it may be observed thatthe results for $k_{F\chi}$=10,
and 20 MeV remain consistent with GW170817 data, while $k_{F\chi}=30\,\mbox{MeV}$ fails to meet this observational constraints. In Fig.\ref{fig:atidal_set3}, we present the same results for a smaller mediator mass.
For set 3 with the hierarchical coupling of the $Z'$ field (left panel), the tidal deformability is larger compared to the case for  set-1 and set-2 with a heavier portal mass. With increase in $k_{F\chi}$, however, the EOS becomes softer with the star becoming more compact and less deformable. On the right panel we show the results for set-4 with a larger coupling of the portal to nucleons. Here, for smaller densities of the DM, the tidal deformability become
larger than pure hadronic matter neutron star.
 As discussed earlier, the vector portal induces a repulsive interaction leading to less compact NSs with DM. This leads to a consistently larger value for the tidal deformability as compared to a pure hadronic matter NSs for lower densities of DM. As $k_{F\chi}$ increases, the EOS starts  becoming softer, leading to larger compactness, and, results in smaller values of tidal deformability.}

\begin{figure}[htb!]
    \centering
    \includegraphics[width=0.48\linewidth]{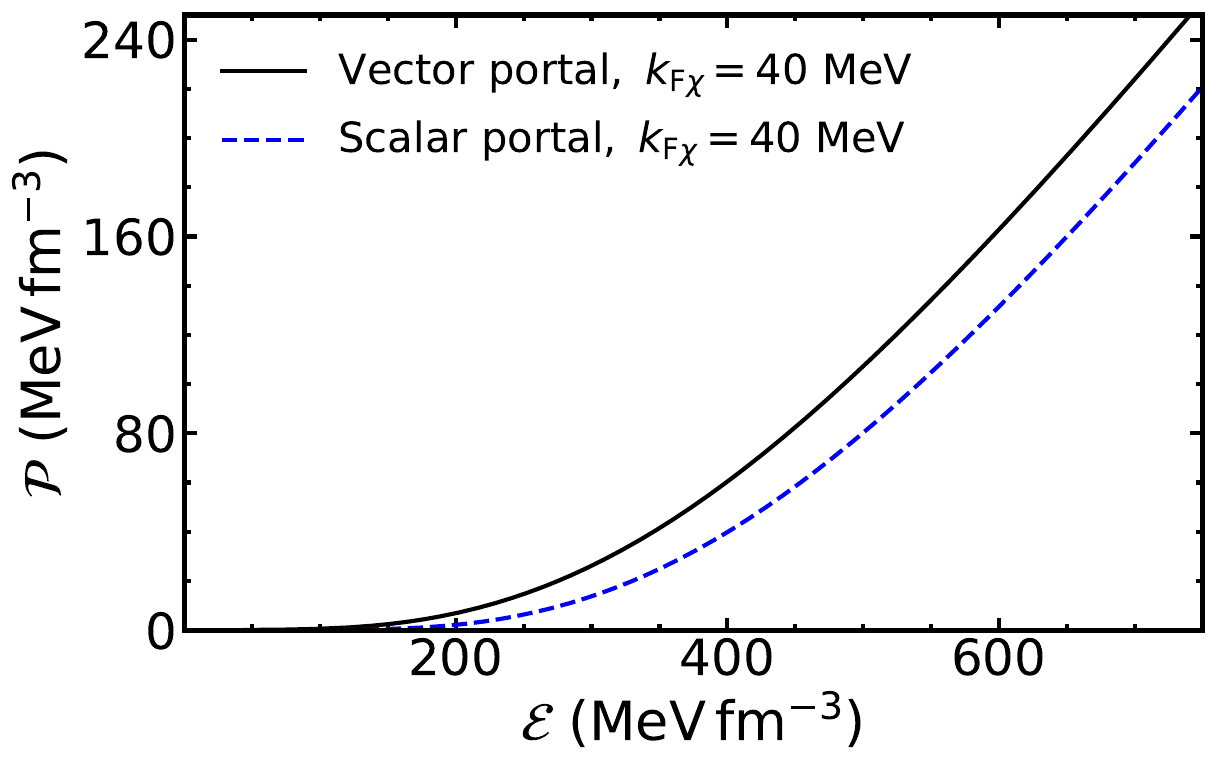}
    \includegraphics[width=0.48\linewidth]{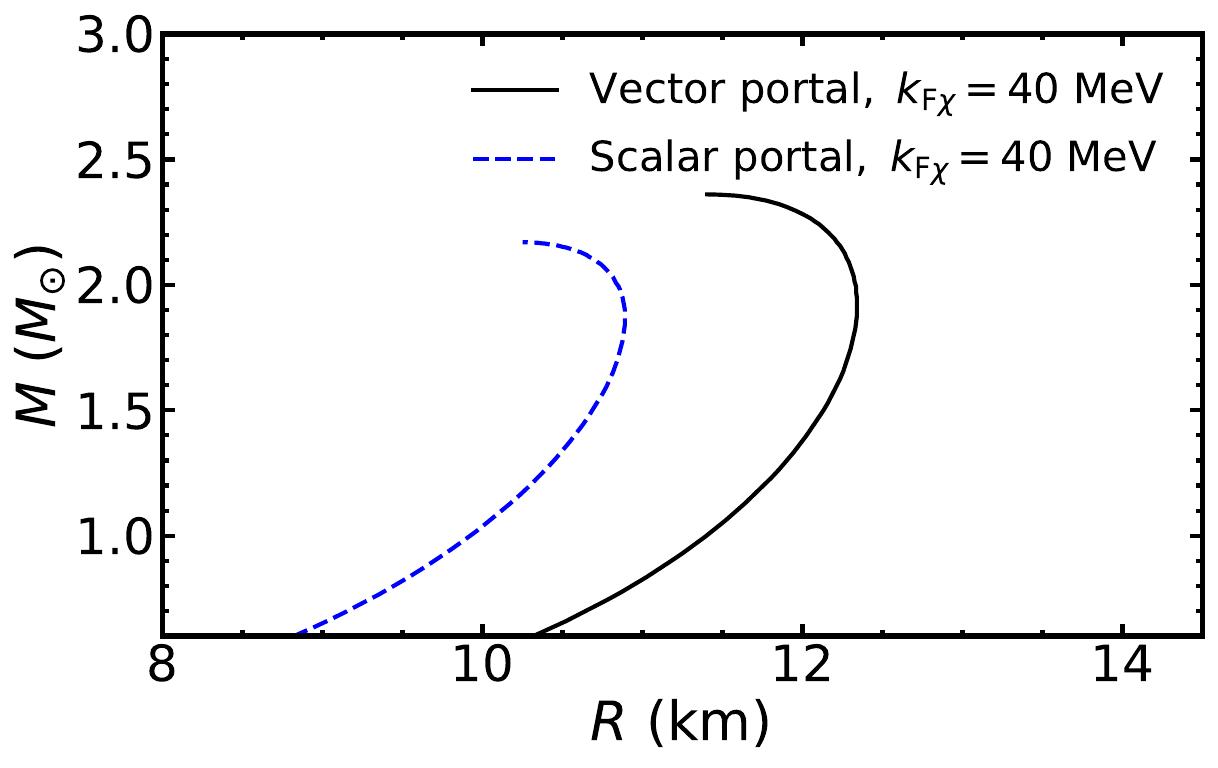}
    \caption{Comparison of NS properties in the presence of fermionic DM interacting via vector and scalar portals for a fixed DM mass $M_{\chi} = 200$ GeV and Fermi momentum $k_{F\chi} = 30$ MeV. The left panel shows the equation of state (pressure $P$ versus energy density $\mathcal{E}$), while the right panel displays the corresponding mass–radius relations obtained from the TOV equations. The dashed line correspond to the Higgs portal with  the parameters taken from \cite{Das:2020ecp, Hajkarim:2024ecp}.}
    \label{fig:slr_vtr_40_EoS_and_amr}
\end{figure}
Finally, it may be worthwhile to compare our mass-radius relations in the present case of vector mediator with that of Higgs portal studies for DM admixed NS matter~\cite{Das:2020ecp, Hajkarim:2024ecp}. 
In Fig.\ref{fig:slr_vtr_40_EoS_and_amr}, we  have compared EOS and the mass-radius relations resulting from the present vector mediated model with the same resulting from the Higgs portal DM model. For the vector portal. we have taken the parameters of set 3 with a lighter vector mediator mass so that the portal contribution to the EOS is significant. It turns out that the maximum mass and maximum radius of the present vector portal model is larger compared to those of Higgs (scalar) mediated models. The origin for such a result lies in stiffening of the EOS compared to the scalar portal model. We might mention here that in this figure, we have taken a larger DM fraction $k_{F\chi} = 40$ ~MeV so as to make the difference between the two scenarios significant. 
\begin{figure}[htb!]
    \centering
    \includegraphics[width=0.48\linewidth]{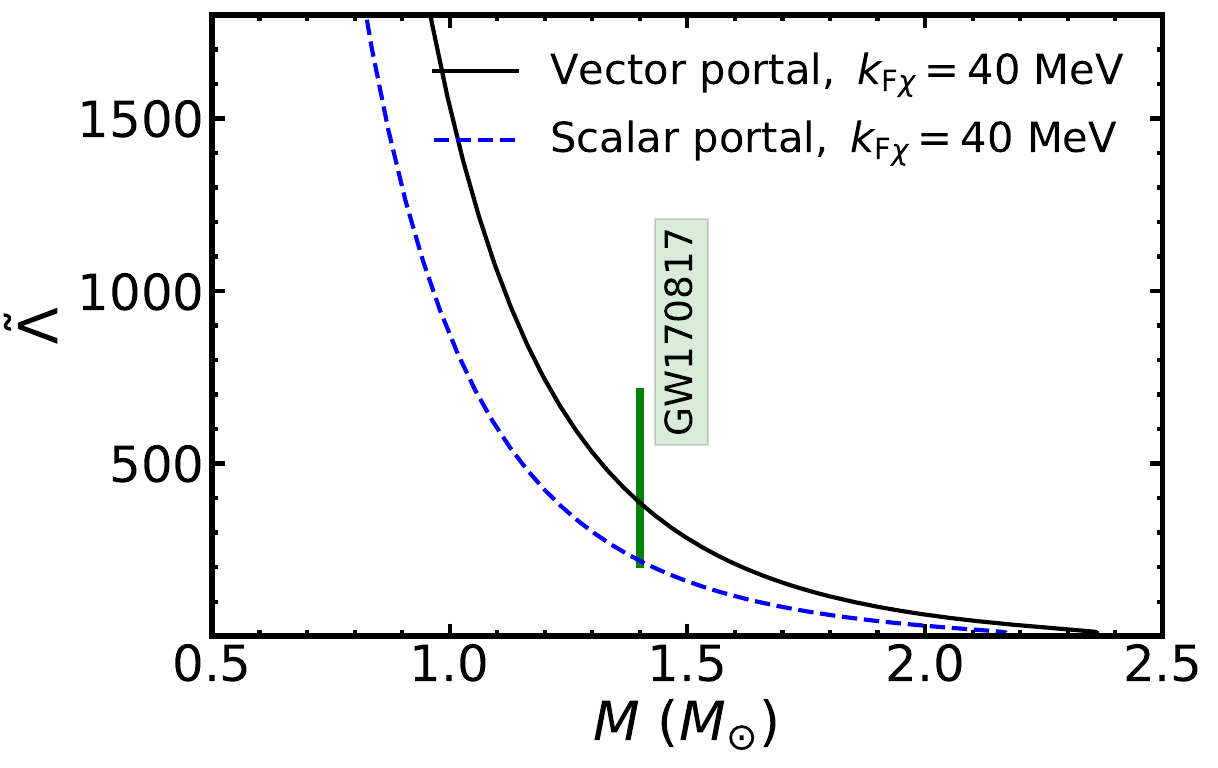}
    \caption{Dimensionless tidal deformability ($\tilde{\Lambda}$) as a function of NS mass for two DM EOS scenarios at fixed Fermi momentum ($k_{{\rm F}\chi} = 40\ \mathrm{MeV}$). The curves compare a vector portal interaction and a scalar portal interaction, highlighting their impact on tidal deformability across the stellar mass range.}
    \label{fig:slr_vtr_40_atidal}
\end{figure}
In Fig.~\ref{fig:slr_vtr_40_atidal}, we compare the dimensionless tidal deformability $\widetilde{\Lambda}$ for the vector portal with set 4 parameters along with the same using the scalar portal. The tidal deformability $\widetilde{\Lambda}$ with the vector portal is consistently larger compared to those resulting from  DM model with a scalar portal. This is again a direct consequence of the repulsive vector interaction, which stiffens the EoS, leading to larger radii and reduced compactness.

\section{Summary and conclusions} \label{sec:summary_and_conclusion}
In this work, we have investigated the impact of fermionic DM interacting with nucleonic matter inside neutron stars through a vector portal ($Z'$). Within the RMF framework, we consistently incorporated contributions from baryons, leptons, DM, and the vector mediator ($Z^\prime$) to construct the EOS of dense matter under conditions of $\beta$-equilibrium and electrical charge neutrality. We find that the presence of the mediator $Z^\prime$  introduces a repulsive interaction that modifies the effective chemical potentials of both nucleons and DM particles. Unlike the often used scalar portal models, where the dominant effect arises through a reduction of the effective nucleon mass leading to  softening of the EOS, the vector portal contributes directly to the pressure via additional vector interactions that affect the effective chemical potential. As a result, EOS can become stiffer depending on the strength of the couplings $g_{\chi Z'},\ g_{q Z'}$ and the mass of the vector mediator $m_{Z^\prime}$ compared to a scalar portal DM scenario. 

A key outcome of our numerical analysis is the clear correlation between DM properties and NS observables such as the mass-radius relation and tidal deformability. For scenarios with heavy mediators [$m_{Z'} \sim \mathcal{O}(100$--$1000)\,$GeV], the vector portal contribution to EOS gets suppressed, and the dominant contribution arises from DM energy density. In this regime, increasing the DM fraction (characterized by larger $k_{F\chi}$) leads to a systematic softening of the EOS, resulting in more compact NSs with a reduction in maximum mass and the corresponding radius which results in a reduction of the tidal deformability. This effect is further amplified for larger DM masses, where the reduction in pressure support leads to significant deviations from the pure nuclear matter case, and in some cases, tensions with observational bounds from GW170817.

The interesting feature of the present study is the distinct role played by light-mediator scenarios in modifying the properties of dark-matter-admixed neutron stars. In particular, the set-3 benchmark is motivated by self-interacting dark matter (SIDM) frameworks involving a light vector mediator and a hierarchical coupling structure, where the mediator couples strongly to the dark sector while maintaining highly suppressed couplings to visible-sector particles. Such scenarios have been extensively studied as viable solutions to several small-scale structure anomalies of collisionless cold dark matter and can simultaneously satisfy current laboratory, direct-detection and collider constraints. Our results show that even within this phenomenologically viable region of parameter space, light vector mediators can produce observable modifications to the equation of state and the global properties of dark-matter-admixed neutron stars. In such case, the portal contribution is non-negligible
resulting in a stiffer EOS leading to a larger tidal deformability compared to a heavy portal mass.
For completeness, we also considered a strong-coupling of a light-mediator 
to both dark and visible sector i.e benchmark set-4 to explore the maximal sensitivity of neutron-star observables to vector portal interactions. While such a benchmark is generally difficult to realize within a minimal kinetically mixed dark-photon framework, it may arise naturally in a more general scenarios involving asymmetric dark matter, secluded dark sectors, or non-minimal gauge extensions. The set-4 benchmark therefore serves primarily as an illustrative example of the largest possible astrophysical effects that can be induced by vector-mediated dark matter interactions. However, we would like to mention that while in the present exploratory investigation we have used the same  RMF parameters, a detailed  Baysian analysis for the parameters of the model in the presence of DM is desirable. 

For light mediator scenarios ($m_{Z'} \sim \mathcal{O}(100)\,$MeV), in particular, the vector portal induces a repulsive interaction that becomes relevant at high densities. In this case, the EOS exhibits a stiffening behavior at higher nuclear densities, leading to larger NS radii and enhanced tidal deformability. This behavior is in sharp contrast with scalar portal models, where the interaction is attractive and usually leads to a softening of the EOS. Consequently, tidal deformability emerges as a particularly sensitive observable that can discriminate between different mechanisms of interaction of the DM. Our results indicate that while heavy mediator scenarios tend to reduce tidal deformability ($\Lambda$), light vector mediators with significant coupling to nuceons can increase it.  For a sufficient fraction of DM, the vector portal with large mass becomes inconsistent with current observational data from gravitational wave observations (GW170817) in LIGO/Virgo~\cite{LIGOScientific:2017vwq, LIGOScientific:2017ync} and X-ray observations of pulsar PSR J0030+0451 in NICER \cite{Vinciguerra:2023qxq}. Thus, the combined analysis of NS observables and terrestrial experiments offers a promising pathway to uncover the nature of DM and its interactions with visible matter under extreme conditions.

\section*{Acknowledgment}
SP acknowledges the financial support under MTR/2023/000687 funded by SERB, Govt. of India. DK also wishes to express his gratitude for the warm hospitality extended to him at Kamala Nibas, Bhubaneswar.

\appendix
\section{Role of vector interactions and equation of state}
In this appendix, we illustrate explicitly how a vector interaction mediated by a $Z'$ boson leads to a repulsive contribution and consequently stiffens the equation of state (EOS). For clarity, we consider a simplified system 
of nucleons interacting via a vector field in the mean-field approximation.

In the mean-field limit, where only the temporal component of the vector field survives, i.e., $\langle Z'_\mu \rangle = Z'_0 \delta_{\mu 0}$, the Hamiltonian density is given by
\begin{equation}
   \mathcal{H}_{\rm MF} = 
   \psi^\dagger \big(-i \boldsymbol{\alpha} \cdot \nabla + \beta M \big) \psi 
   + g_{N Z'} \psi^\dagger \psi \, Z'_0 
   - \frac{1}{2} m^2_{Z'} Z_0'^2 \, .
\end{equation}
The thermodynamic potential at zero temperature is defined as
\begin{equation}
    \Omega = \langle \mathcal{H}_{\rm MF} \rangle - \mu_B n_B,
    \label{eq:Omega_app}
\end{equation}
where $\mu_B$ is the baryonic chemical potential and $n_B$ is the number density given by
\begin{equation}
n_B = \frac{\gamma k_F^3}{6\pi^2},
\end{equation}
with $\gamma=2$ for spin degeneracy. The expectation value of the Hamiltonian density becomes
\begin{equation}
\langle \mathcal{H}_{\rm MF} \rangle =
\frac{\gamma}{(2\pi)^3} \int_0^{k_F} d^3k \, \sqrt{k^2 + M^2}
+ g_{N Z'} n_B Z'_0
- \frac{1}{2} m^2_{Z'} Z_0'^2.
\end{equation}
The effective chemical potential is shifted due to the vector mean field as $\mu_B^* = \mu_B - g_{N Z'} Z'_0$, which determines the Fermi momentum through $k_F = \sqrt{\mu_B^{*2} - M^2}$. Minimizing the thermodynamic potential with respect to the mean field,
\begin{equation}
\frac{\partial \Omega}{\partial Z'_0} = 0
\quad \implies 
Z'_0 = \frac{g_{N Z'}}{m_{Z'}^2} n_B.
\label{eq:Zp_solution}
\end{equation}

Substituting Eq.~(\ref{eq:Zp_solution}) back into the Hamiltonian, the total energy density can be written as
\begin{equation}
\mathcal{E} =
\frac{1}{\pi^2} (3\pi^2 n_B)^{4/3} \, E_N(M/k_F)
+ \frac{g_{N Z'}^2}{2 m_{Z'}^2} n_B^2,
\end{equation}
where the function $E_N(x)$ encodes the relativistic kinetic contribution,
\begin{equation}
E_N(x) =
\frac{1}{8} \left[
\sqrt{1+x^2} (2+x^2)
- x^4 \ln\left(\frac{1+\sqrt{1+x^2}}{|x|}\right)
\right].
\end{equation}

The pressure is obtained from the thermodynamic relation
\begin{equation}
\mathcal{P} = -\Omega = \mu_B n_B - \mathcal{E},
\end{equation}
which yields
\begin{equation}
\mathcal{P} =
\frac{1}{3\pi^2} (3\pi^2 n_B)^{4/3} \, P_N(M/k_F)
+ \frac{g_{N Z'}^2}{2 m_{Z'}^2} n_B^2,
\label{eq:P_final}
\end{equation}
with
\begin{equation}
P_N(x) =
\frac{1}{8} \left[
\sqrt{1+x^2} (2-3x^2)
+ 3x^4 \ln\left(\frac{1+\sqrt{1+x^2}}{|x|}\right)
\right].
\end{equation}
Thus, it receives an additional repulsive contribution that grows rapidly with density. A key observation is that the vector interaction contributes a term proportional to $n_B^2$ to both the energy density and pressure,
\begin{equation}
\mathcal{E}_{Z'} = \mathcal{P}_{Z'} =
\frac{g_{N Z'}^2}{2 m_{Z'}^2} n_B^2,
\end{equation}
which is manifestly positive and therefore repulsive in nature.

\begin{figure}
    \centering
    \includegraphics[width=0.7\linewidth]{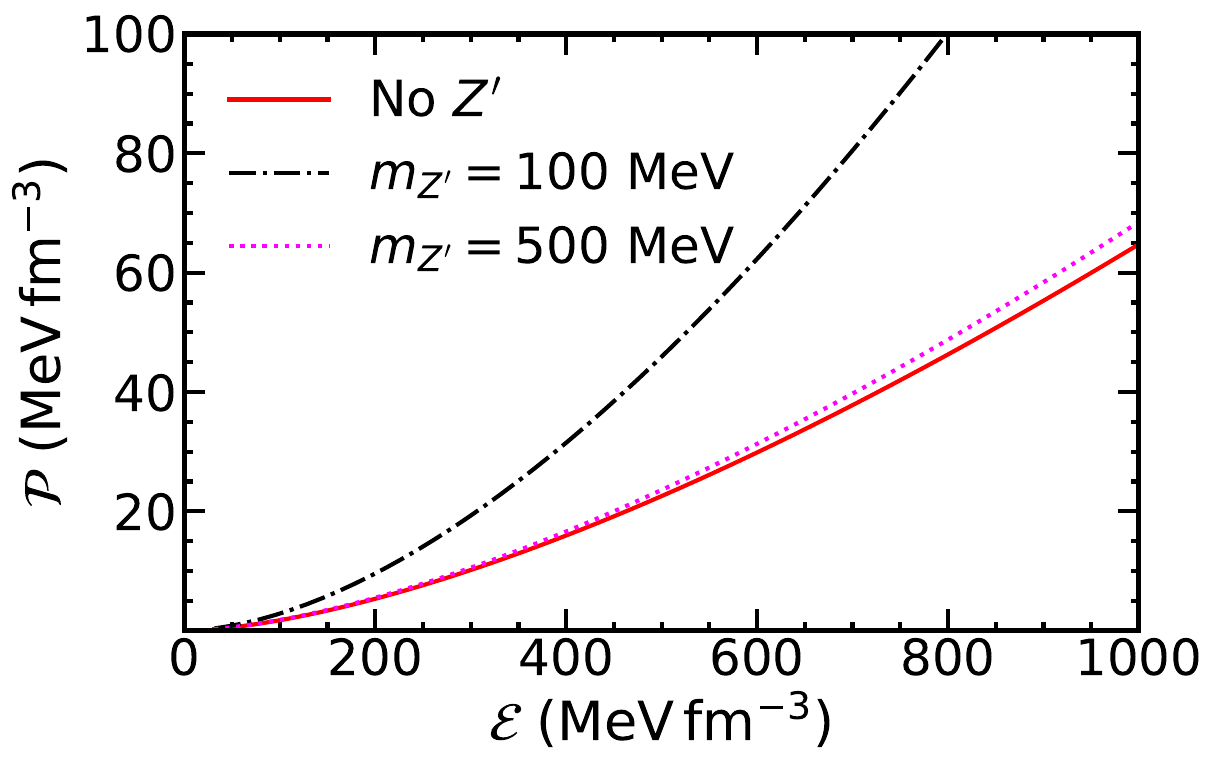}
    \caption{
Effect of the vector portal interaction mediated by $Z^\prime$ on the equation of state, showing pressure as a function of energy density for different values of the mediator mass $m_{Z'}$. The black solid curve corresponds to the case without $Z'$ interaction, while the colored curves represent finite vector portal contributions with $m_{Z'} = 100$ MeV and $500$ MeV. The inclusion of the $Z'$ mediator leads to an enhancement of pressure at a given energy density, with the effect becoming stronger for smaller mediator masses. This shows the  repulsive nature of the vector interaction and thereby, the equation of state becomes progressively stiffer, particularly at high densities relevant for NS cores.}
\label{fig:zprime_effect}
\end{figure}

The total energy density and pressure can be expressed schematically as
\begin{eqnarray}
\mathcal{E} = A_{\mathcal{E}} n_B^{4/3} + B n_B^2, \qquad
\mathcal{P} = A_{\mathcal{P}} n_B^{4/3} + B n_B^2, \mbox{with}\, 
B = \frac{g_{N Z'}^2}{2 m_{Z'}^2}.
\end{eqnarray}
The $n_B^{4/3}$ term arises from the kinetic contribution, while the $n_B^2$ term originates from the vector interaction. At sufficiently high densities, the quadratic term dominates, leading to a rapid increase of pressure with density.

The baryonic chemical potential is given by
\begin{equation}
\mu_B = \frac{\partial \mathcal{E}}{\partial n_B}
= \mu_B^{(0)} + \frac{g_{N Z'}^2}{m_{Z'}^2} n_B,
\end{equation}
which clearly shows that the vector interaction increases the energy cost of compression linearly with density, a hallmark of repulsive interactions.

In realistic NS matter, nucleons interact via scalar ($\sigma$), vector ($\omega$), and isovector ($\rho$) meson fields within the RMF framework. The $\sigma$ field provides an attractive interaction by reducing the effective nucleon mass, thereby softening the EoS, while the $\omega$ field introduces a repulsive contribution proportional to $n_B^2$, similar to the vector interaction derived above. The inclusion of a $Z'$-mediated vector portal interaction extends this framework by introducing an additional repulsive channel between baryons and DM. Its contribution, also scaling as $n_B^2$, directly enhances the pressure at high densities. For heavy mediator masses, this contribution is suppressed and the EOS is primarily governed by standard RMF interactions and DM. However, for light mediators, the $Z'$ term can become comparable to the $\omega$-meson contribution, leading to significant stiffening of the EOS as displayed in Fig.\ref{fig:zprime_effect}.

\bibliographystyle{jhep}
\bibliography{zzbib}
\end{document}